\documentclass[lineno]{jfm}

\usepackage{graphicx}
\usepackage{epstopdf,epsfig}
\usepackage{newtxtext}
\usepackage{newtxmath}
\usepackage{natbib}
\usepackage{algorithm}
\usepackage{algpseudocode}
\usepackage{caption}
\usepackage{subcaption}
\usepackage{hyperref}
\hypersetup{
    colorlinks = true,
    urlcolor   = blue,
    citecolor  = black,
}

\newcommand{\RomanNumeralCaps}[1]

\title{Stochastic Wavevector Model for Rapidly-Distorted Compressible Turbulence}

\author{Noah Zambrano\aff{1}
  \corresp{\email{nzamb@umich.edu}}\and 
  Karthik Duraisamy\aff{1}
 }

\affiliation{\aff{1}Department of Aerospace Engineering, University of Michigan, Ann Arbor, MI 48103
}

\begin{document}
\maketitle
\begin{abstract}
A stochastic wavevector approach is formulated to accurately represent compressible turbulence subject to rapid deformations. This approach is inspired by the incompressible particle representation model of \citet{kassinos1995structure} and preserves the exact nature of compressible Rapid Distortion Theory (RDT). The adoption of a stochastic - rather than the Fourier - perspective simplifies the transformation of statistics to physical space and serves as a starting point for the development of practical turbulence models. We assume small density fluctuations and isentropic flow to obtain a transport equation for the pressure fluctuation. This results in four fewer transport equations compared to the compressible RDT model of \citet{yu2007extension}. The final formulation is closed in spectral space and only requires numerical approximation for the transformation integrals. The use of Monte Carlo for unit wavevector integration motivates the representation of the moments as stochastic variables. Consistency between the Fourier and stochastic representation is demonstrated by showing equivalency between the evolution equations for the velocity spectrum tensor in both representations. Sample clustering with respect to orientation allows for different techniques to be used for the wavevector magnitude integration. The performance of the stochastic model is evaluated for axially-compressed turbulence, serving as a simplified model for shock-turbulence interaction, and is compared to LIA and DNS. Pure and compressed sheared turbulence at different distortion Mach numbers are also computed and compared to RDT/DNS data. Finally, two additional deformations are applied and compared to solenoidal and pressure-released limits to demonstrate the modeling capability for generic rapid deformations.
\end{abstract}

\section{Introduction}
\label{sec:Introduction}
A majority of practical turbulence models rely on the Boussinesq approach which relates the Reynolds stress anisotropy $\overline{u'_i u_j'}-q^2\delta_{ij}/3$, where $q^2$ is twice the turbulent kinetic energy, directly to the mean rate-of-strain $S_{ij}$ through the eddy-viscosity $\nu_t$, such that the Reynolds stress is determined by the rate of strain. This is generally a reasonable approximation at low mean flow deformation rates because turbulence has time to respond to the mean flow. However, the eddy-viscosity assumption fails to accurately represent turbulence at large or rapid deformations rates $S \tau_t>>1$, where $S$ is the (large) shear strain rate and $\tau_t$ is the turbulent time scale. In these flows, the turbulence reacts to rapid deformations at a rate different from the mean-flow time scales and behaves more like an elastic solid 
 \citep{crow1968viscoelastic,taylor1935turbulence}. Thus, the viscous fluid-like qualitative behavior that an eddy viscosity assumption imposes \citep{crow1968viscoelastic} is no longer accurate. For large deformations, rapid distortion theory (RDT) provides an accurate representation of the turbulence by linearizing the governing equations. RDT, however, is unclosed in inhomogeneous turbulent flows due to the non-locality associated with the fluctuating pressure. While most relevant engineering applications are inhomogeneous, it is still valuable to develop models for simplified homogeneous problems and extend it to inhomogeneous cases. The advantage of building a model beginning from homogeneous RDT is that it requires no modeling assumptions. The motivation for this work is to develop such a model from homogeneous RDT specifically for compressible turbulence such that a future, more complete model can be developed with a rigorous foundation.

In contrast to developing exact solutions and analyzing the behavior of compressible turbulence, as was done in many past compressible RDT studies \citep{durbin1992rapid,simone1997effect,jacquin1993turbulence,goldstein1978unsteady}, we develop a modeling framework that pursues RDT analysis from a stochastic perspective, rather than a spectral one.The stochastic designation is different from most common stochastic methods which include random perturbation terms in the model for turbulent evolution \citep{pope2001turbulent}. The formulation presented in this work has deterministic evolution equations that evolve a stochastic initial state, as a first step towards a more general modeling framework. A stochastic perspective allows the use of stochastic integration methods for transforming spectral/stochastic statistics to physical space. This technique was pioneered by \citet{kassinos1995structure} for incompressible turbulence, where more details on the equivalency between the spectral and stochastic perspective is given in the book by \cite{sagaut2008homogeneous}. The most recent work to extend this idea to compressible turbulence was done by \citet{yu2007extension}, \citet{bertsch2012rapid}, and \citet{lavin2012flow}, but requires solving 21 coupled transport equations.
 
 RDT is valid when the timescale of mean distortion rate is much shorter than the timescale of turbulent interactions \citep{batchelor1954effect}. This limiting case is achieved by considering very large deformation values in the mean deformation tensor $\partial \bar{U}_i/\partial x_j$. The advantage to RDT is that it simplifies the governing equations by linearizing the fluctuation equations with respect to the mean flow \citep{durbin2011statistical}. Fortunately, for homogeneous turbulence, this method is exact in that no modeling assumptions are made. Linearization is accomplished by neglecting higher-order fluctuating terms in the fluctuating equations. The linearized first-moment equations are then used to formulate evolution equations for the second-moments. For compressible RDT, the usage of a Helmholtz decomposition and transformation in Craya-Herring coordinates is popular because significant simplifications can be made based on the symmetries of the mean flow or properties of the fluctuating field. In this frame of reference, the dilatational part of the velocity field is aligned with the wavevector and two orthogonal solenoidal components lie in the plane normal to the wavevector \citep{jacquin1993turbulence,cambon1993rapid,simone1997effect}. Rather than using this decomposition and transformation, we only apply the transformation for deforming coordinates \citep{rogallo1981numerical}, and then solve for fluctuating quantities in this reference frame. A similar aspect to the Craya-herring frame where the dilatational part of the velocity field is aligned with the wavevector while the solenoidal part is orthogonal is also realized in deforming coordinates. In addition, the  method of randomly initializing isotropic turbulence in \citet{rogallo1981numerical} is interpretable using the Craya-Herring frame. The deforming coordinate transformation is effectively a transformation from a Eulerian to a Lagrangian perspective and thus gives the basis for a particle representation of turbulence. Solenoidal and dilatational components for the second-moments are found after-the-fact by using both the  compressible and solenoidal transport equations for turbulence statistics.

An important parameter to characterize rapidly strained compressible turbulence is the distortion Mach number $M_d = S\ell/a$ \citep{durbin1992rapid},  where $\ell$ is an integral length-scale and $a=\sqrt{\gamma \bar{p}/\bar{\rho}}$ is the mean speed of sound. The distortion Mach number gives a quantitative sense of which regime the turbulence lies in. \citet{jacquin1993turbulence} and \citet{cambon1993rapid} defined the solenoidal regime for low $M_d$ values and the pressure-released regime for higher $M_d$ values. In these regimes, either the influence of dilatational effects are ignored (solenoidal) or the pressure effects are considered negligible (pressure-released). When $M_d$ is near these limits, the governing equations greatly simplify such that analytical solutions are possible or the computation of exact solutions is straight-forward. The distortion Mach number is used to study the performance of the new stochastic model across a range bounded by the limits and can be compared to DNS/RDT solutions of past studies. Observing how the model predictions change across $M_d$ will also display the unique behavior of compressible turbulence described in past studies.

In essence, the primary focus of this work is to extend the particle representation model (PRM) of \cite{kassinos1995structure} to compressible turbulence. One of the first wavevector models for turbulence appears in the thesis by \cite{fung1990kinematic}, where the Eulerian velocity field is represented by a large number of Fourier coefficients and the wavevector orientation is chosen randomly. This approach is also called a \textit{Kinematic Simulation} and has been extended to anisotropic cases by \cite{cambon2004turbulent} and \cite{favier2010space}. A similar anisotropic stochastic wavevector model is described  by \cite{kassinos1995structure}, where the initial conditions are randomly generated (hence the stochastic part), and evolved in time deterministically. By shifting to a Lagrangian perspective, we can treat the instantaneous variables as stochastic realizations or ``particles" and numerically solve the transformation integrals. In compressible flows, however, the fluctuating velocity field is not solenoidal which causes the wavevector magnitude to appear in the evolution equation for the velocity spectrum tensor and prevents closure at the directional velocity spectrum tensor level. This issue is addressed by introducing an additional integration step but requires more particles compared  to the directional velocity spectrum integration method. Our model is also an improved version of the compressible RDT model developed by \cite{yu2007extension}. The new model is able to achieve better results using four fewer transport equations. This is accomplished by defining a new normalized pressure fluctuation variable that enables closure using isentropic relations. Additionally, to demonstrate consistency between the stochastic and Fourier formulations, we show that integration with respect to the wavevector state space of the stochastic velocity spectrum tensor is equivalent to the integration of the velocity spectrum with respect to the Fourier modes. We also address the inconsistency of using the directional spectrum integration and the unnecessary use of the full velocity spectrum tensor rather than just the symmetric part. Finally, the new compressible PRM model is tested on various cases and compared to RDT/DNS data. Axially-compressed turbulence is verified against the DNS data of \citet{cambon1993rapid} and used as a simplified shock-turbulence model for varying initial Mach numbers. This simple shock-turbulence model is similar to the idea proposed by \citet{jacquin1993turbulence} and is compared to post-shock predictions with linear interaction approximations (LIA). Performance in homogeneous shear without any mean compression is also evaluated against the DNS/RDT data of \citet{simone1997effect} and provides a basis for comparison against the RDT model of \citet{yu2007extension}. The solenoidal version of the model is verified for a sheared compression case against the RDT data of \citet{mahesh1996interaction}. Finally, application to generic rapid mean deformations is demonstrated by applying the model to plane strain and axisymmetric contraction and verifying the limiting solenoidal/pressure-released behavior for each deformation.

\section{Governing equations and basic assumptions}
\label{sec:Governing equations}
This work  develops the evolution equations for homogeneous anisotropic compressible turbulence using the Reynolds decomposition, which states that $u=\bar{u}+u'$ such that $\bar{u}$ is the ensemble average (also called Reynolds average) and $u'$ is the fluctuation from the mean. The fluctuation field in compressible turbulence includes solenoidal/dilatational velocity modes and an entropy mode. The dilatational velocity is coupled to pressure fluctuations such that it satisfies an acoustic wave equation \citep{simone1997effect}. In this work, we consider compressible flows where the fluctuation field is composed of both modes (typically called compressible turbulence). In compressible flows, the mean velocity field is not divergence free, i.e. $\nabla\cdot \bar{\mathbf{u}}\neq0$. This condition arises from the continuity equation when density varies. The fluctuation field is not divergence free either, i.e. $\nabla\cdot\mathbf{u}'\neq0$. This means that the fluctuating velocity field has both solenoidal and dilatational components, where $\nabla\cdot\mathbf{u}_s'=0$ and $\mathbf{u}_s$ denotes the solenoidal component and $\mathbf{u}_d$ the dilatational component.

The restriction of homogeneity on compressible flows is more stringent as compared to the incompressible case. Both homogeneous incompressible and compressible turbulence require that the mean velocity gradient has the form $\partial \bar{u}_i/\partial x_j=A_{ij}(t)$. However, as shown by \cite{blaisdell1991numerical}, the mean density and pressure are only functions of time because the right-hand side of the evolution equation for the density and pressure statistics must also only be functions of time \citep{pope2001turbulent}. This limitation is only true when the density dilatation $\overline{\rho\partial u_i'/\partial x_i}$ and pressure dilatation  $\overline{p\partial u_i'/\partial x_i}$ are not zero \citep{blaisdell1991numerical}, which is true for the compressible flows considered. Another assumption used to simplify the governing equation is that of small density fluctuations compared to the mean density. While some variable-density flows may have non-negligible density fluctuations compared to the mean (as described by \cite{sandoval1995dynamics}), this work primarily focuses on turbulence where the ratio  $\rho'/\bar{\rho}$ is approximately zero.

The governing equations comprise of the invsicid conservation of momentum, continuity, and conservation of fluctuating entropy. The equations are
\begin{equation}
    \frac{\partial \rho}{\partial t}=-\frac{\partial}{\partial x_i}(\rho u_i),
\end{equation}
\begin{equation}
    \frac{\partial \rho u_i}{\partial t}=-\frac{\partial}{\partial x_j}(\rho u_iu_j+p\delta_{ij}),
\end{equation}
\begin{equation}
    \frac{\partial s'}{\partial t}=-u_i\frac{\partial s'}{\partial x_i},
\end{equation}
where $s'$ is the specific entropy fluctuation. If we assume an adiabatic process for an ideal gas with constant specific heat, then the conservation of entropy and the isentropic relation $p/\rho^{\gamma}=\textrm{constant}$ implies that $\gamma D(\rho'/\bar{\rho})/Dt=D(p'/\bar{p})/Dt$, where, $D/Dt$ is the substantial derivative. By using the continuity equation and the former relation to relate the pressure fluctuation to the velocity fluctuation divergence $\partial u_i'/\partial x_i$, the variables required to close the second-moment equations are considerably reduced when compared to the use of the equation of state and conservation of energy as done by \cite{yu2007extension}.This technique was used by other authors in studies of compressible turbulence as well \citep{durbin1992rapid,simone1997effect,cambon1993rapid}.

\section {Representing compressible rapid distortions}
RDT analysis is used to develop  closed-form evolution equations for relevant moments of interest. The linear solutions are valid for turbulent flows with distortions that are sufficiently rapid such that $S\tau_t = S\ell/q>>1$. Applicability is also limited to short times and small wavenumber modulus $\kappa$, such that $\kappa q/S<<1$.

The evolution equations for the velocity and pressure fluctuations are the only equations required to close the Reynolds stress evolution equation. The first step is the linearization of the governing fluctuation equations. Full fluctuation equations are obtained by traditional methods as described by \cite{wilcox1998turbulence}. Higher-order fluctuating terms are then removed to yield the linearized equations:
\begin{equation}\label{eq:continuity}\frac{\partial \rho'}{\partial t}+\bar{u}_j\frac{\partial \rho'}{\partial x_j}=-\bar{\rho}\frac{\partial u_j'}{\partial x_j}-\rho'\frac{\partial \bar{u}_j}{\partial x_j},\end{equation}

\begin{equation}
\label{eq:momentum}\frac{\partial u_i'}{\partial t}+\bar{u}_j\frac{\partial u_i'}{\partial x_j}=-u_j'\frac{\partial \bar{u}_i}{\partial x_j}-\frac{1}{\bar{\rho}}\frac{\partial p'}{\partial x_i},\end{equation}

\begin{equation}\frac{\partial}{\partial t}\left(\frac{p'}{\gamma \bar{p}}\right)+\bar{u}_j\frac{\partial}{\partial x_j}\left(\frac{p'}{\gamma \bar{p}}\right)=-\frac{\partial u'_i}{\partial x_i}.\end{equation}

For purely solenoidal  modes, pressure fluctuation influences are decoupled and equations simplify with the use of a fluctuating pressure Poisson equation derived from equation \ref{eq:momentum} and the divergence free condition $\partial u'_i/\partial x_i =0$,
\begin{equation}\frac{1}{\bar{\rho}}\frac{\partial^2 p'}{\partial x_i^2}=-2\frac{\partial u_j'}{\partial x_i}\frac{\partial \bar{u}_i}{\partial x_j}-\bar{u}_j\frac{\partial^2 u_j'}{\partial x_i^2}.\end{equation}

The pressure Poisson equation will be used to derive the solenoidal component of the compressible RDT of turbulence. Next, a Fourier transform is applied and the reference frame is changed from a Eulerian to Lagrangian point of view. This is done by introducing the substantial derivative $D/Dt$ and is equivalent to transforming into deforming coordinates as done by \cite{rogallo1981numerical},
\begin{equation}
\label{eq:fluc-vel}\frac{D\hat{u}_i}{Dt}=-\frac{\partial \bar{u}_i}{\partial x_k}\hat{u}_k+\mathrm{i}\kappa_i\frac{\hat{p}}{\bar{\rho}},\end{equation}
\begin{equation}\label{eq:fluc-k}
\frac{D\kappa_i}{Dt}=-\kappa_k\frac{\partial \bar{u}_k}{\partial x_i},\end{equation}
\begin{equation}\label{eq:fluc-pres}
\frac{D}{Dt}\left(\frac{\hat{p}}{\gamma \bar{p}}\right) =\mathrm{i}\kappa_k\hat{u}_k.\end{equation}
The hat notation denotes a Fourier-transformed variable and $\mathrm{i}$ is the imaginary number. $\kappa_i$ is the component of the wavevector, where the evolution equation in homogeneous turbulence from a Lagrangian point of view is obtained from \cite{kassinos1995structure}. Again, these RDT first-moment equations are similar to  those used in other compressible RDT studies  \citep{durbin1992rapid,simone1997effect,cambon1993rapid}. For solenoidal modes, the pressure Poisson equation in wavespace is
\begin{equation}\label{eq:pressure_poisson}
\frac{-\kappa_i^2\hat{p}}{\bar{\rho}}=2\mathrm{i}\kappa_i\hat{u}^s_j\frac{\partial \bar{u}_i}{\partial x_j}.\end{equation}
The evolution of the one-point Reynolds stress $R_{ij}=\overline{u'_iu'_j}$ in physical space is the target of this analysis. The one-point Reynolds stress is related to the two-point Reynolds stress correlation $R_{ij}(\mathbf{r})=\overline{u'_i(\mathbf{x})u'_j(\mathbf{x}')}$, where \citet{durbin2011statistical} defines the two-point correlation through the symmetric part of the velocity spectrum tensor,
\begin{equation}\label{eq:rs_def}
\overline{u_i'(\mathbf{x})u_j'(\mathbf{x}')} =\iiint^{\infty}_{-\infty}\Phi_{ij}(\boldsymbol{\kappa}) e^{\mathrm{i}\boldsymbol{\kappa}\cdot (\mathbf{x}'-\mathbf{x})}d^3\boldsymbol{\kappa}.\end{equation}
When $\mathbf{x}'=\mathbf{x}$, then the two-point correlation reduces to the one-point correlation $R_{ij}$ which is then equal to the integral of $\Phi_{ij}$ across all $\boldsymbol{\kappa}$.

\subsection{Normalized pressure fluctuations and second moments}
A new variable, the normalized pressure fluctuation (denoted as $\hat{\wp}=\mathrm{i}\hat{p}/(\gamma\bar{p})$ henceforth) is introduced to simplify the upcoming derivations for the second-moments and to remove dependence on imaginary terms. The evolution equation for $\hat{\wp}$ follows equation \ref{eq:fluc-pres},
\begin{equation}\label{eq:fluc-wp}
\frac{D\hat{\wp}}{Dt} =-\kappa_k\hat{u}_k.\end{equation}
Next, the second-moments must be determined. The velocity spectrum tensor $\Phi_{ij}$ is used to find the one-point Reynolds stress $R_{ij}$, but its evolution equation is unclosed. We now define the single-point pressure-velocity spectrum vector ($\beta_i$) and pressure spectrum scalar ($\Gamma$) to help close the $\Phi_{ij}$ evolution equation. These additional spectra are defined in a similar manner to the velocity spectrum tensor.  The symmetric spectra are defined as
\begin{equation}\label{eq:vel-spec-ten}
\Phi_{ij} =\frac{1}{2} \left(\overline{ \hat{u}_i\hat{u}^+_j}+\overline{ \hat{u}^+_i\hat{u}_j}\right),\end{equation}
\begin{equation} \beta_{i}=\frac{1}{2}\left(\overline{\hat{\wp}^+\hat{u}_i} + \overline{\hat{\wp}\hat{u}_i^+}\right),\end{equation}
\begin{equation} \Gamma=\overline{\hat{\wp}^+\hat{\wp}},\end{equation}
where the $+$ superscript represents the complex conjugate. The velocity spectrum tensor $\Phi_{ij}$ is defined as the real components of the ensemble average of the Fourier-transformed velocity fluctuations \citep{durbin2011statistical}. The final first-moment equations are then established by substituting in the normalized pressure fluctuation, 
\begin{equation}
\label{eq:fm_vel}\frac{D\hat{u}_i}{Dt}=-\frac{\partial \bar{u}_i}{\partial x_k}\hat{u}_k+\kappa_ia^2\hat{\wp},\end{equation}
\begin{equation}\label{eq:fm_k}
\frac{D\kappa_i}{Dt}=-\kappa_k\frac{\partial \bar{u}_k}{\partial x_i},\end{equation}
\begin{equation}\label{eq:fm_pres}
\frac{D\hat{\wp}}{Dt} =-\kappa_k\hat{u}_k.
\end{equation}
The appearance of the $\kappa_ia^2\hat{\wp}$ term is associated with the dispersion frequency of acoustic waves \citep{sagaut2008homogeneous}. This introduces new difficulties with the integration of spectral higher-order moments to physical space. With the first moment evolution equations established, the second-moment equations are derived by substituting in the time-derivatives of the first-moments and differentiating with respect to time using the product rule,
\begin{equation} \label{eq:velSpecTenEvo}
\frac{D}{Dt}(\Phi_{ij})=-\frac{\partial \bar{u}_i}{\partial x_k}\Phi_{kj}-\frac{\partial \bar{u}_j}{\partial x_k}\Phi_{ki}+\kappa_ia^2\beta_j+\kappa_ja^2\beta_i,\end{equation}
\begin{equation}\frac{D}{Dt}(\beta_i)=-\kappa_k\Phi_{ki}-\frac{\partial \bar{u}_i}{\partial x_k}\beta_{k}+\kappa_ia^2\Gamma,\end{equation}
\begin{equation}\label{eq:presSpecEvo}\frac{D}{Dt}(\Gamma)=-2\kappa_i\beta_i. \end{equation}
These final spectral second-moment equations are fully closed and require no new modeling assumption aside from the assumptions made in RDT.

\subsection{Transformation to physical space}
\label{sec:transform}
The final step is to take an inverse Fourier transform to obtain statistics in physical space. This is an integration over all wave number components, which is simplified by transforming into spherical coordinates. For example, the evolution equation for the Reynolds stress becomes
\begin{equation}\label{eq:spherical-trans}
\frac{D}{Dt}(R_{ij})=\int^{2\pi}_0 \int^\pi_0 \int^{\infty}_{-\infty} \left(-\frac{\partial \bar{u}_i}{\partial x_k}\Phi_{kj}-\frac{\partial \bar{u}_j}{\partial x_k}\Phi_{ki}+\kappa_ia^2\beta_j+\kappa_ja^2\beta_i\right) \kappa^2\sin{\theta}d\kappa d\theta d\phi\end{equation}
where $\kappa=\sqrt{\kappa_i\kappa_i}$ and $\kappa_i$ can be decomposed into product of wavevector magnitude $\kappa$ and orientation $\theta,\phi$. Note that the appearance of $\kappa_i$ means we cannot analytically integrate with respect to the wavevector magnitude as done by \cite{kassinos1995structure}. The solenoidal velocity spectrum tensor, however, can be integrated this way. The latter expression can be solved using a fast-Fourier transform to get the most accurate results. However, for this work, we chose to approach the integration through  stochastic methods and numerical integration. 


\subsection{Higher-order tensors}
The transformation into Fourier space allows for relatively easier  computation of several higher-order tensors. These tensors are used to obtain more information about the structure of turbulence. The structure tensors as defined by \citet{kassinos1996particle} are of particular interest. The dimensionality tensor is based on the kinetic energy spectrum,
\begin{equation}
D_{ij}=\iiint^{\infty}_{-\infty}\frac{\kappa_i\kappa_j}{\kappa^2}\Phi_{mm}d^3\boldsymbol{\kappa},
\end{equation}
and gives information on whether the turbulence is one- or two-dimensional. The dimensionality gives information on the type of eddies present in the flow. The eddy types are the one-dimensional jettal eddy, the two-dimensional vorticial eddy, and the helical eddy \citep{kassinos1995structure}. The structure tensor information is also required to accurately represent rotating flows \citep{reynolds1989effects}.  Similarly, the circulicity tensor, which describes the vorticial field, is also found easily in Fourier space,
\begin{equation}
F_{ij}=\epsilon_{inm}\epsilon_{jts}\iiint^{\infty}_{-\infty}\frac{\kappa_n\kappa_t}{\kappa^2}\Phi_{ms}d^3\boldsymbol{\kappa}.
\end{equation}
In the former equation, $\epsilon_{ijk}$ is the alternating tensor. Finally, higher-rank tensors are also found without requiring additional transport equations. An important fourth-rank tensor,
\begin{equation}
M_{ijpq}=\iiint^{\infty}_{-\infty}\frac{\kappa_p\kappa_q}{\kappa^2}\Phi_{ij}d^3\boldsymbol{\kappa}
\end{equation}
gives information regarding the fluctuating pressure field and is often used to represent non-local information as it contains information on two-point statistics \citep{kassinos1995structure}. 
 
\subsection{Limitations} 
The three main assumptions of the current formulation imply limitation to a certain subset of flows. The most substantial assumptions are those of linearization about a base flow, small density fluctuations compared to the mean, and the isentropic assumption. Linearization is valid only for turbulence subject to rapid deformation. This simplification allows the closure of the equations for the first and second moments without any phenomenological modeling. However, it implicitly prohibits the application of the formulations to slower deformations, as experienced in many engineering flows of interest. We take linearization as a first step towards more general models, where it can be viewed as a limiting behavior with the formulation matching predictions from RDT when subject to increasingly large deformations. however, the formulation is expected to gradually become imprecise as deformation decreases.

To represent slower deformations, additional  modeling approximations must be added to the formulation and will be explored in upcoming work. The assumption of small density fluctuations also causes discrepancies with application to problems with large compression ratios. The rationale is that as the mean density changes, the magnitude of the fluctuations can become significant. While we do assume that the density fluctuation is small compared to the mean, it is \textit{weakly enforced}. The simplification is used in the derivation of equations \ref{eq:continuity} and \ref{eq:momentum}, where it is assumed that terms scaled by $\rho'/\bar{\rho}$ are small enough to be ignored when compared in magnitude to the other terms. However, $\rho'/\bar{\rho}$ is not assumed to be zero when considering its rate of change. The isentropic relation uses $D(\rho'/\bar{\rho})/Dt$ to derive the pressure equation, and thus the influence of the density fluctuation with respect to the mean is accounted for in the pressure terms. This weak assumption was similarly applied by \cite{jacquin1993turbulence} and \cite{cambon1993rapid} in their studies of compressed homogeneous turbulence.

Finally, the isentropic relation limits the current formulation to adiabatic and entropy-conserving flows. While many compressible flows have non-entropy conserving features such as shocks, we choose to invoke isentropy due to the significant reduction in equations. If isentropy was not invoked, the pressure equation would need to be derived in a similar manner to \cite{yu2007extension}, where the ideal gas law and energy equation are used to relate temperature and density to pressure evolution. This would result in 4 more transport equations; two first-rank tensors and two scalars. This amounts to 8 more transport equations, which adds to the computational cost. This work seeks a simple formulation which is applicable to most regions where compressible turbulence is important. Non-isentropic parts of the flow  usually require additional treatment. For example, turbulence models for flows with shocks require special shock-capturing methods or variable Prandtl number \textit{ad hoc} modifications \cite{roy2018variable}  to accurately predict the shock, and non-adiabatic flows such as hypersonic boundary layers require special treatment to account for the inhomogeneity from the wall. We believe that the treatment for various issues that appear in non-isentropic flows can be combined with the required modifications to the current formulation, rather than incorporated in the base formulation.

\section{Stochastic Modeling}
In a stochastic representation, flow variables are treated  as  {\color{red}random fields}. Since the Fourier transform is a linear operation, the Fourier transformed random fields are also random fields in $\boldsymbol{\kappa}$-space. Formally, the random field can be written as $\phi(x,t)=\int  e^{-i\boldsymbol{\kappa}\cdot\mathbf{x}}dZ(\boldsymbol{\kappa}, t)$, where $Z(\boldsymbol{\kappa}, t )$ is a complex-valued random measure with the property that $\mathbb{E}[dZ(\boldsymbol{\kappa}, t)] = 0$ and $\mathbb{E}[dZ(\boldsymbol{\kappa}_1, t_1)dZ^*(\boldsymbol{\kappa}_2, t_2)] = \delta(\boldsymbol{\kappa}_1 - \boldsymbol{\kappa}_2)\delta(t_1 - t_2)S(\boldsymbol{\kappa})d\boldsymbol{\kappa}_1d\boldsymbol{\kappa}_2dt_1dt_2$, where $S(\boldsymbol{\kappa})$ is the power spectral
density. From these properties, the spectral representation of the stochastic fields have the following mode properties: $\mathbb{E}[\hat{\phi}(\boldsymbol{\kappa}, t)] = 0$ and $\mathbb{E}[\hat{\phi}(\boldsymbol{\kappa}_1, t_1)\hat{\phi}^*(\boldsymbol{\kappa}_2, t_2)] = \delta(\boldsymbol{\kappa}_1 - \boldsymbol{\kappa}_2)\delta(t_1 - t_2)S(\boldsymbol{\kappa})$, which match the properties of the physical spectral modes. This is the primary mathematical rationale for why the properties of Fourier modes allows for representation as stochastic realizations. 
Section~\ref{sec:particle_properties} details the specific flow-variables associated with each stochastic representation. From a physical perspective, this is supported by the hypothesis that in homogeneous turbulence, velocity fluctuations at points separated by large distances tend to become uncorrelated. The de-correlation in wavevector space for widely different wavevectors also enforces the idea of a turbulence energy cascade which suggests that energy is transferred predominantly between nearby scales.

Another motivation for shifting to a stochastic representation is the ease of transformation from spectral to physical space. As discussed in section \ref{sec:transform}, the Fourier modes must be integrated with respect to all wavevectors. One way of doing this integration is through a Monte Carlo integral. Monte Carlo integration for some general function $f(\mathbf{x})$ over a domain $\Omega$ is given by \cite{caflisch1998monte},
\begin{equation}\int_\Omega f(\mathbf{x})d \mathbf{x} \approx \frac{V}{N_s}\sum^{N_s}_{i=1}f(\mathbf{x}_i),\; \;V=\int_\Omega d\mathbf{x},\end{equation}
where $N_s$ is the number of samples that are uniformly sampled from $\Omega$. This integration approximation approaches the expected value $\mathbb{E}(f(\mathbf{x}))$ as $n\rightarrow \infty$. An important concern regarding the accuracy of the integration is realized by noting the infinite integration bounds of equation \ref{eq:spherical-trans}. This must be approximated where some cut-off wavevector magnitude is specified \textit{a priori}. As done in the previously, we use equation \ref{eq:spherical-trans} to convert to a spherical integral and explicitly express the wavevector in terms of orientation and magnitude. In the work by \cite{kassinos1995structure}, the infinite integration with respect to magnitude was done analytically as the wavevector magnitude did not appear in the evolution equation for the conditional Reynolds stress, thus avoiding the problem of approximating the infinite integral bounds. This meant that the incompressible PRM evolved the directional velocity spectrum tensor, 
\begin{equation} \Phi^{\nearrow \kappa}_{ij}(\theta,\phi) =\int^{\infty}_0 \kappa^2 \Phi_{ij}(\kappa,\theta,\phi) d\kappa,\end{equation}
instead of the velocity spectrum tensor. This is the lowest level at which the incompressible formulation can be closed \citep{van1997pdf}. \cite{kassinos1995structure} also scaled the directional velocity spectrum tensor by a factor of $4\pi$ at initialization rather than at each timestep. This constant appears in the Monte Carlo integration with respect to orientation. By evolving this quantity and averaging across all samples, \citet{kassinos1995structure} obtained an equivalent method to the Monte Carlo integration with respect to the wavevector orientation,
\begin{equation}R_{ij}=\int^{2\pi}_0 \int^\pi_0 \Phi^{\nearrow \kappa}_{ij} \sin{\theta}d\theta d\phi\approx \frac{1}{N_s}\sum^{N_s}_{n=1}4\pi \Phi^{\nearrow \kappa}_{ij}(\theta^{(n)},\phi^{(n)}),\end{equation}
where the $(n)$ superscript denotes the sample number. Integrating with respect to all wavevector magnitudes is advantageous because it reduces the number of samples required to effectively and accurately use Monte Carlo integration. This is so because it avoids integration of the energy spectrum. The step-like shape of the energy spectrum is not well-suited for Monte Carlo integration, where the requirement of uniform sampling makes it difficult to resolve the sharp feature and dramatically increases sample count. Unfortunately, integrating analytically with respect to all wavevector magnitudes is not possible for the compressible case as the time derivative of the velocity spectrum tensor is a function of the wavevector orientation and magnitude, whereas in the incompressible/solenoidal case it is independent. We must introduce another Monte Carlo approximation in order to obtain a solution. This involves selecting a range of wavevector magnitudes $\Delta k$, such that a Monte Carlo integral is used to find the directional velocity spectrum tensor at each particle
\begin{equation}\Phi^{\nearrow \kappa}_{ij}(\theta^{(n)},\phi^{(n)})\approx\frac{\Delta k}{N_{sh}}\sum^{N_{sh}}_{{sh}=1}(\kappa^{(sh)})^2\Phi_{ij}(\kappa^{(sh)},\theta^{(n)},\phi^{(n)}),\end{equation}
where the $sh$ sub/superscript represents the ``shell" sample of the wavevector magnitude. To simplify the methodology, we can combine the two Monte Carlo integrals into one, where we sample from a set of particles that each have a unique wavevector magnitude and orientation. This particle realization differs from the cluster method developed by \cite{kassinos1995structure} as we no longer group particles based on orientation and rather perform the integration for each wavevector component in one large Monte Carlo integration
\begin{equation}R_{ij}\approx\frac{4\pi \Delta k}{N_s}\sum^{N_s}_{n=1}(\kappa^{(n)})^2\Phi_{ij}(\boldsymbol{\kappa}^{(n)}).\end{equation}
The insight from this analysis shows that PRM is identical to RDT for every step up until the integration. PRM uses an approximate  Monte Carlo (and thus stochastic) integration technique rather than a Fourier transform algorithm to go from the closed set of equations in Fourier space to physical space.

\subsection{Stochastic particle properties and identities}\label{sec:particle_properties}
To complete the Monte Carlo integration, the velocity spectrum tensor must be sampled. This provides the motivation for representing the velocity spectrum as a stochastic variable from which we treat samples as Lagrangian particles. We call these stochastic samples particles to follow the terminology of \citet{kassinos1995structure}. 

\begin{figure}
\centerline{\includegraphics[width=0.7\textwidth]{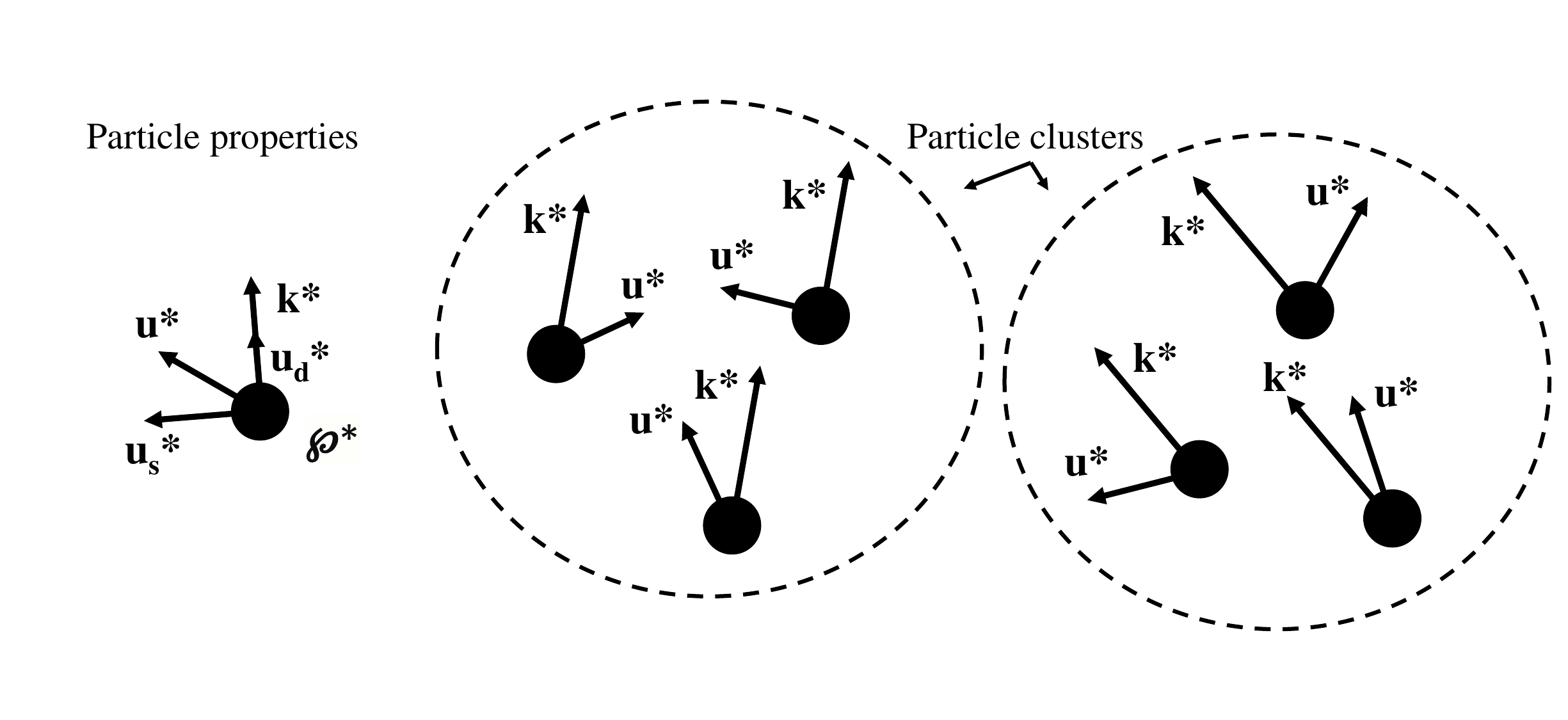}}
  \caption{Particle properties associated with a cluster of samples.}
\label{fig:particle_properties}
\end{figure}


Under the assumptions previously stated, the stochastic representation is valid for the turbulence considered in this study. We can now formally define the stochastic variables. Suppose there exist multivariate distributions in $\boldsymbol{\kappa}$-space for $\Phi_{ij}(\boldsymbol{\kappa},t), \beta_i(\boldsymbol{\kappa},t)$, and $\Gamma(\boldsymbol{\kappa},t)$  such that $\mathcal{K} \subset \mathbb{R}^3$ is the state space of the wavevector. We can obtain samples from $\boldsymbol{\kappa}^{(1)},\boldsymbol{\kappa}^{(2)},...,\boldsymbol{\kappa}^{(n)} \in \mathcal{K},$
Where the $(n)$ superscript denotes sample number. The second moments defined previously correspond to realizations of the underlying distributions. Now, we sample (for a range of $\boldsymbol{\kappa}$ vectors) from the initial random distributions for the three second-moment probability density functions (PDF) and advance their solutions forward in time using the evolution equations for the second-moments. Note that the stochastic behavior comes from the initialization rather than from the evolution equations. Each realization for $\Phi_{ij}, \beta_i$, $\kappa_i$, and $\Gamma$ is evolved in time separately. 

Particles are assigned properties which are evolved through time and are used to compute second-moment statistics. The assigned properties are the particle velocity vector $\mathbf{u}^*$ (which can be decomposed into solenoidal and dilatational components), the particle wavevector $\boldsymbol{\kappa}^*$, and the particle normalized pressure $\wp^*$. These stochastic particles represent a one-dimensional one-component flow, a basic building block for the turbulence structure. The particle velocity vector, wavevector, and normalized pressure can be thought of as stochastic samples of the respective Fourier modes. Note that compared to incompressible PRM, the particle wavevector and velocity vector are not orthogonal since the velocity field is not divergence free. However, the solenoidal velocity vector $\mathbf{u}^{*s}$ is orthogonal to $\boldsymbol{\kappa}^*$ and the dilatational velocity vector $\mathbf{u}^{*d}$ is parallel to $\boldsymbol{\kappa}^*$ such that $u_i^{*s}\kappa_i=0$ and $u^*_i=u^{*s}_i+u^{*d}_i$. This is illustrated in figure \ref{fig:particle_properties}. In addition, the particle wavevector is similar but not identical to the gradient vector defined by \citet{kassinos1995structure}. Since the stochastic particles are analogous to the Fourier modes of the flow variables, the stochastic governing equations are also analogous to the RDT equations. The first-moment equations are
\begin{equation}\label{eq:fm_stochastic_vel}
\frac{Du^*_i}{Dt}=-\frac{\partial \bar{u}_i}{\partial x_k}u^*_k+\kappa^*_ia^2\wp^*,\end{equation}
\begin{equation}\label{eq:fm_stochastic_wv}
\frac{d\kappa^*_i}{Dt}=-\kappa^*_k\frac{\partial \bar{u}_k}{\partial x_i},\end{equation}
\begin{equation}\label{eq:fm_stochastic_pres}
\frac{D\wp^*}{Dt} =-\kappa^*_ku^*_k.\end{equation}
The closed-set of second-moment stochastic equations are
\begin{equation} \label{eq:sm_stochastic_vel}
\frac{D}{Dt}(\Phi^*_{ij})=-\frac{\partial \bar{u}_i}{\partial x_k}\Phi^*_{kj}-\frac{\partial \bar{u}_j}{\partial x_k}\Phi^*_{ki}+\kappa^*_ia^2\beta^*_j+\kappa^*_ja^2\beta^*_i, \end{equation}
\begin{equation}\label{eq:sm_stochastic_vel_pres}
\frac{D}{Dt}(\beta^*_i)=-\kappa^*_k\Phi^*_{ki}-\frac{\partial \bar{u}_i}{\partial x_k}\beta^*_{k}+\kappa^*_ia^2\Gamma^*,\end{equation}
\begin{equation}\label{eq:sm_stochastic_pres}
\frac{D}{Dt}(\Gamma^*)= -2\kappa^*_i\beta^*_i. \end{equation}

\subsection{Equivalency with spectral representation}
This section presents a more rigorous justification to show equivalency with spectral and stochastic representations. We turn to the PDF formulation of compressible RDT. While PDF methods are typically used to model the one-point, one-time Navier-Stokes equations \citep{pope2001turbulent}, they also provide a more grounded justification for using a stochastic point of view for turbulence.  In this section, the marginal PDF is used to relate the random stochastic variables to the Fourier spectra. To show equivalency and consistency between the PDF and traditional formulations, the stochastic velocity spectrum tensor must have the same evolution equation form as the Fourier velocity spectrum,
\begin{equation}\label{eq:equivalency_req}
    \overline{u'_iu'_j}=\iiint^{\infty}_{-\infty}\Phi_{ij}(\boldsymbol{\kappa})d^3\boldsymbol{\kappa}\equiv\iiint^{\infty}_{-\infty}\Phi^*_{ij}(\boldsymbol{\mathcal{K}})d^3\boldsymbol{\mathcal{K}}=\overline{u^*_iu^*_j}.
\end{equation}
The stochastic velocity spectrum is defined similarly to  \citet{van1997pdf}, where it is equal to the integral of the product of the velocity state space and joint
PDF $f^*(\mathbf{V},\boldsymbol{\mathcal{K}})$. However, since the velocity also depends on the normalized pressure fluctuation $\wp$, the full state space of the joint PDF contains a total of nine variables (two vectors and a scalar), $f^*(\mathbf{V},\boldsymbol{\mathcal{K}},\mathcal{P})$, for which the normalized pressure state space $\mathcal{P}$ must first be integrated out to give the joint PDF dependent only on velocity and wavevector states. The stochastic pressure-velocity and pressure spectra are defined in a similar fashion. The stochastic second-order moments are
\begin{equation}\label{eq:stochastic_vel_spec}
    \Phi^*_{ij}(\boldsymbol{\mathcal{K}})=\iiiint^{\infty}_{-\infty} V_iV_jf^*(\mathbf{V},\boldsymbol{\mathcal{K}},\mathcal{P})d\mathcal{P}d^3\mathbf{V}=\langle u_i^*u_j^*|\boldsymbol{\kappa}^*=\boldsymbol{\mathcal{K}}\rangle f^*(\boldsymbol{\mathcal{K}}),
\end{equation}
\begin{equation}\label{eq:stochastic_vel_pres_spec}
    \beta^*_{i}(\boldsymbol{\mathcal{K}})=\iiiint^{\infty}_{-\infty} V_i\mathcal{P}f^*(\mathbf{V},\boldsymbol{\mathcal{K}},\mathcal{P})d\mathcal{P} d^3\mathbf{V}=\langle u_i^*\wp^*|\boldsymbol{\kappa}^*=\boldsymbol{\mathcal{K}}\rangle f^*(\boldsymbol{\mathcal{K}}).
\end{equation}
\begin{equation}\label{eq:stochastic_pres_spec}
    \Gamma^*_{i}(\boldsymbol{\mathcal{K}})=\iiiint^{\infty}_{-\infty} \mathcal{P}^2f^*(\mathbf{V},\boldsymbol{\mathcal{K}},\mathcal{P})d\mathcal{P}d^3\mathbf{V} =\langle \wp^*\wp^*|\boldsymbol{\kappa}^*=\boldsymbol{\mathcal{K}}\rangle f^*(\boldsymbol{\mathcal{K}}).
\end{equation}
The joint PDF $f^*(\mathbf{V},\boldsymbol{\mathcal{K}},\mathcal{P})$ is derived using traditional methods \citep{heinz2003statistical} and is based on the stochastic first-moment equations (equations \ref{eq:fm_stochastic_vel}-\ref{eq:fm_stochastic_pres}).
\begin{equation}\label{eq:marginal_PDF}
    \frac{\partial f^*}{\partial t}=\frac{\partial \bar{u}_k}{\partial x_n}\frac{\partial}{\partial \mathcal{K}_n}(\mathcal{K}_kf^*)+\frac{\partial \bar{u}_n}{\partial x_k}\frac{\partial}{\partial V_n}(V_kf^*)+\frac{\partial}{\partial V_n}(a^2 \mathcal{K}_n \mathcal{P} f^*)-\frac{\partial}{\partial \mathcal{P}}(\mathcal{K}_kV_kf^*)
\end{equation}
Since a Eulerian PDF is used, the stochastic velocity spectrum evolution equation must match the non-conservative Eulerian velocity spectrum evolution equation (transformed from equation \ref{eq:velSpecTenEvo}),
\begin{equation}\label{eq:eulerian_vel_spec}
    \frac{\partial \Phi_{ij}}{\partial t}=\kappa_m\frac{\partial \bar{u}_m}{\partial x_n}\frac{\partial \Phi_{ij}}{\partial \kappa_n}+\frac{\partial \bar{u}_k}{\partial x_k}\Phi_{ij}-\frac{\partial \bar{u}_j}{\partial x_k}\Phi_{ki}-\frac{\partial \bar{u}_i}{\partial x_k}\Phi_{kj}-\kappa_ia^2\beta_j-\kappa_ja^2\beta_i.
\end{equation}
Now, the derivative of equation \ref{eq:stochastic_vel_spec} with respect to time is taken to find the stochastic evolution equation. Note that the state space variables $\mathbf{V}, \mathcal{P},\boldsymbol{\mathcal{K}}$ are constant in time, but the random variables sampled from them are not. Substituting in equation \ref{eq:marginal_PDF} into the time derivative of equation \ref{eq:stochastic_vel_spec} and integrating each term yields the stochastic velocity spectrum transport equation. Integration by parts is used along with linearity to simplify the solution for each term. The first two terms only depend on integrating with respect to $\mathbf{V}$ since no $\mathcal{P}$ terms appear,
\begin{equation}\label{eq:term1}
   \iiint^{\infty}_{-\infty} V_iV_j\frac{\partial \bar{u}_k}{\partial x_n}\frac{\partial}{\partial \mathcal{K}_n}(\mathcal{K}_kf^*)d^3\mathbf{V}=\frac{\partial \bar{u}_k}{\partial x_n}\frac{\partial \Phi^*_{ij}}{\partial \mathcal{K}_n}\mathcal{K}_k+\frac{\partial \bar{u}_k}{\partial x_k}\Phi^*_{ij},
\end{equation}
\begin{equation}\label{eq:term2}
    \iiint^{\infty}_{-\infty} V_iV_j\frac{\partial \bar{u}_n}{\partial x_k}\frac{\partial}{\partial V_n}(V_kf^*)d^3\mathbf{V}=-\frac{\partial \bar{u}_i}{\partial x_k}\Phi^*_{kj}-\frac{\partial \bar{u}_j}{\partial x_k}\Phi^*_{ki}.
\end{equation}
The third term requires four integration steps. Integration is first carried out on $\mathcal{P}$ and then on $\mathbf{V}$. Terms are simplified using the stochastic second-moment definitions (equations \ref{eq:stochastic_vel_spec}-\ref{eq:stochastic_pres_spec}),
\begin{multline}
    \label{eq:term3}
    \iiiint^{\infty}_{-\infty} V_iV_j\frac{\partial}{\partial V_n}(a^2 \mathcal{K}_n \mathcal{P} f^*)d\mathcal{P} d^3\mathbf{V}\\
    =V_iV_ja^2 \mathcal{K}_n \mathcal{P} f^*-\iiiint^{\infty}_{-\infty} (\delta_{ni}V_j+\delta_{nj}V_i)a^2 \mathcal{K}_n \mathcal{P} f^*d\mathcal{P} d^3\mathbf{V}\\
    =-a^2 \mathcal{K}_i \beta^*_j -a^2 \mathcal{K}_j\beta^*_i.
\end{multline}
Finally, the fourth term also requires four integration steps. Due to linearity and independence of $\mathcal{P}$ and $\mathbf{V}$, the last term is zero.
\begin{multline}\label{eq:term4}
    \iiiint^{\infty}_{-\infty} -V_iV_j\frac{\partial}{\partial \mathcal{P}}(\mathcal{K}_kV_kf^*)d\mathcal{P} d^3\mathbf{V}\\
    =-\iiint^{\infty}_{-\infty} V_iV_j\mathcal{K}_kV_kf^*d^3\mathbf{V}+\iiiint^{\infty}_{-\infty} \frac{\partial}{\partial \mathcal{P}}(V_iV_j) \mathcal{K}_kV_kf^*d\mathcal{P} d^3\mathbf{V}=0
\end{multline}
Putting solutions \ref{eq:term1}-\ref{eq:term4} together gives the evolution equation for the stochastic velocity spectrum tensor from an Eulerian frame of reference,
\begin{equation} \label{eq:stochastic_vel_spec_evo}
  \frac{\partial \Phi^*_{ij}}{\partial t}=\mathcal{K}_m\frac{\partial \bar{u}_m}{\partial x_n}\frac{\partial \Phi^*_{ij}}{\partial \mathcal{K}_n}+\frac{\partial \bar{u}_k}{\partial x_k}\Phi^*_{ij}-\frac{\partial \bar{u}_j}{\partial x_k}\Phi^*_{ki}-\frac{\partial \bar{u}_i}{\partial x_k}\Phi^*_{kj}-\mathcal{K}_ia^2\beta^*_j-\mathcal{K}_ja^2\beta^*_i.
\end{equation}
Equations \ref{eq:stochastic_vel_spec_evo} and \ref{eq:eulerian_vel_spec} show that the two equations are identical. Since the two are identical, the stochastic representation is consistent with the Fourier representation and thus gives the same results when computing turbulent statistics in physical space.

\subsection{Particle Clusters}
The integration with respect to wavevector magnitude presents additional difficulties. 
Consider a type of particle \textit{cluster} where particles are grouped together such that they all have the same orientation but different wavevector magnitudes and spectra, as shown in figure \ref{fig:particle_properties}. 
Our idea of clusters is nearly identical to \citet{kassinos1995structure}, but rather than possessing random directional spectra, the clusters have a specified range of wavevector magnitudes associated with it, which directly changes the spectra. 
The motivation for clustering is to reduce the number of samples required to compute accurate statistics and separate the particle information on wavevector magnitude and orientation. Clustering also enables the use of non-stochastic methods for the inner integral,
\begin{equation}\label{eq:approx_int}
R_{ij}\approx \frac{1}{N_c}\sum^{N_c}_{n=1}4\pi \int_0^{\infty} (\kappa^{(sh))})^2\Phi_{ij}(\kappa^{(sh)},\theta^{(c)},\phi^{(c)})d\kappa,\end{equation}
where the $c$ superscript denotes the cluster index. Generally, the wavevector magnitude evolution depends on the wavevector orientation. However, because of the wide magnitude integration limits, the evolution of the wavevector magnitude sometimes plays a negligible role. Therefore, by splitting up the integration steps, we can explicity ignore the magnitude evolution and only need to consider the evolution of the wavevector orientations. The evolution equation for the unit wavevector $\kappa_i/k=e_i$ is
\begin{equation}
\frac{De_i}{Dt}=-e_k\frac{\partial \bar{u}_k}{\partial x_i}+\frac{\partial \bar{u}_k}{\partial x_r}e_ke_re_i.\end{equation}
By only considering the orientation evolution, we are in essence resampling the stochastic particles at each time-step such that they fall within the predefined wavevector magnitude range.  The integral in equation \ref{eq:approx_int} can be computed in various ways, such as trapezoidal, quadrature, etc. In this work, we use the trapezoidal and Gauss-Laguerre quadrature integration methods. Note that this resampling technique only works for some cases. For example, it was found that the pure shear case requires the evolution magnitude for stability reasons.

\section{Application to compressible homogeneous turbulence}
\label{sec:Application to homogeneous flows}
In this section we present results for various types of homogeneous flows. We compare against DNS and RDT data for axial compression, pure shear, and sheared compression. The model is then extended to plane strain and axisymmetric contraction and the limiting behavior is evaluated. Initialization of the stochastic particles is also discussed.

\subsection{Initialization} \label{sec:initialization}
The initialization method will change depending on whether the velocity spectrum tensor or directional velocity spectrum tensor is being evolved. The simpler initialization is the velocity spectrum tensor. Turbulence is initialized from homogeneous isotropic turbulence (HIT), unless otherwise stated. The definition for the velocity spectrum tensor in HIT is given by \cite{durbin2011statistical} in terms of the energy spectrum,
\begin{equation}\label{eq:init-sol}
\Phi_{ij}(\boldsymbol{\kappa}^{(n)})=\frac{E(\kappa^{(n)})}{4\pi (\kappa^{(n)})^2}\left(\delta_{ij}-\frac{\kappa^{(n)}_i\kappa^{(n)}_j}{(\kappa^{(n)})^2}\right),\end{equation}
where $n$ denotes the particle number and $E(K)$ is the energy spectrum. Note that is initialization is only valid for solenoidal turbulence. The velocity spectrum tensor is assigned to each $n^{\textrm{th}}$ particle given its wavevector orientation and magnitude. The initial energy spectrum is specified based on desired initial conditions. In the most general cases, the Von K\'arm\'an energy spectrum is used. It is given as \citep{durbin2011statistical}
\begin{equation}
    E(\kappa)=q^2LC_{vK}\frac{(\kappa L)^4}{[1+(\kappa L)^2]^p},
\end{equation}
where $L$ is the length scale required for dimensional consistency, $q^2=R_{ii}$, and $p=17/6$ to match the -5/3 law. $C_{vK}$ is found by enforcing
\begin{equation}\label{eq:energy_spec_cond}
\int_{-\infty}^{\infty}E(k)d\kappa=\frac{1}{2}q^2,\end{equation}
such that $C_{vK}\approx0.4843$. When comparing results to RDT/DNS data from \cite{simone1997effect} or \cite{cambon1993rapid}, we used the same initial energy spectrum defined in those papers,
\begin{equation}
    E(\kappa)=q^2A\kappa^4\exp{\frac{-2\kappa^2}{\kappa_p^2}},
\end{equation}
where $\kappa_p=8$ is the wavevector magnitude where the energy spectrum peaks, and $A$ is a constant such that equation \ref{eq:energy_spec_cond} is satisfied. To evolve the velocity spectrum tensor, the unit wavevector is initialized randomly about a unit sphere for a specified range of $k$ magnitudes. The number of unique orientations is equal to the cluster sample count, while the number of $\kappa$ discretizations is based on the integration method (logarithmic spacing with $\Delta \kappa$ total range for trapezoidal, for example). To initialize the directional velocity spectrum tensor, only the unit vectors in a unit sphere must be initialized. This is also done randomly. Each unit vector has a unique directional velocity spectrum tensor value such that
\begin{equation}\frac{1}{N}\sum^N_{c=1}4\pi \Phi^{\nearrow \kappa}_{kk}(\mathbf{e}^{(c)})=\frac{1}{2}q^2.\end{equation}
This means that the diagonal values of the Reynolds stress are equal to $\frac{1}{3}q^2$. The HIT initialization for the directional velocity spectrum tensor is then
\begin{equation}4\pi \Phi^{\nearrow \kappa}_{ij}(\boldsymbol{\kappa}^{(c)}/\kappa)=\left(\delta_{ij}-\frac{\kappa^{(c)}_i\kappa^{(c)}_j}{\kappa^2}\right)\int^\infty_{-\infty}E(\kappa^{(c)})d\kappa.\end{equation}
The energy spectrum integral is numerically evaluated using midpoint integration approximation.

Initialization of the pressure-velocity and pressure spectra will require knowing their respective energy spectra. Generally, these are not known. For the examples in this work, the velocity-pressure spectrum is set to zero for HIT. However, following the initialization of the solenoidal/dilatational modes in \cite{simone1997effect}, the strong form of acoustic equilibrium \citep{sarkar1991analysis} is enforced when the pressure spectrum is defined by the dilatational component of the energy spectrum. The ratio of solenoidal to dilatational turbulent kinetic energy $q^2_s/q^2_d$ must first be specified and used to scale the energy spectrum $E(k)$, such that $E_s(\kappa)+E_d(k)=E(\kappa)$ (where $E_s(\kappa)/E_d(\kappa)=q^2_s/q^2_d$) , to obtain the following initialization:
\begin{equation}\Gamma(\boldsymbol{\kappa}^{(n)})=\frac{E_d(\kappa^{(n)})}{2\pi (\kappa^{(n)})^2a_0^2}.\end{equation}
The influence of solenoidal and dilatational effects depend on the distortion Mach number $M_d$ \citep{durbin1992rapid}. For each case, and initial distortion Mach number $M_d=M_tSq^2/\epsilon$ is specified along with a turbulent Mach number $M_t=q/a$ and turbulent Reynolds number $Re_{t}=q^4/(\nu\epsilon)$. These three conditions along with the relationship between the total turbulent dissipation rate and energy spectrum,
\begin{equation}
    \epsilon=2\nu \int_0^{\infty}\kappa^2E(\kappa)d\kappa,
\end{equation}
are used to initialize the stochastic spectra. Finally, it is noted that for the pure shear and shear compression case, $\gamma=1.4$ is used whereas $\gamma=5/3$ is used for the axisymmetric compression. All cases use a constant initial mean temperature of $\bar{T}=300$.The temperature is not assumed to remain constant throughout the simulation, and its change is captured through the change in pressure and speed of sound. The density and temperature fluctuations do not explicitly appear in the evolution equations and thus are not required to be explicitly initialized. However, they can be initialized through the isentropic relations between density, temperature, and pressure.

\subsection{Algorithm for stochastic representation}
All the background necessary for a functional algorithm to evolve the turbulent statistics of compressible turbulence has been discussed in the prior sections. The complete algorithm is now presented in the following pseudo code to further solidify methodology. Depending on the case, the following variables are known/chosen: the final non-dimensional time $St_f$, the number of particle clusters $n_{p}$, the number of wavevector modulus discretizations $n_k$, the number of time steps $n_t$,the initial mean speed of sound $a_0$, the initial turbulent Mach number $M_{t0}$, the initial distortion Mach number $M_{d0}$, the initial turbulent Reynolds number $Re_{t0}$, the initial solenoidal-to-dilatational turbulent kinetic energy ratio $q_s^2/q_d^2$, the initial shear-to-compression ratio $S_0/C_0$, and the maximum and minimum wavevector magnitudes $\kappa_{\textrm{min}}, \kappa_{\textrm{max}}$. The governing equations are solved numerically using finite time discretization to advance the solution forward in time. We used an explicit fourth-order Runge-Kutta scheme.

\begin{algorithm} [H]
\caption{Compressible Particle Representation Model}\label{alg:CPRM}
\begin{algorithmic}
\State Inputs: $St_f,n_{p},n_k,n_t,a_0, M_{t0}, M_{d0}, Re_{t0}, q_s^2/q_d^2, S_0/C_0, \kappa_{\textrm{min}}, \kappa_{\textrm{max}}$
\State Compute $q^2_0, \Delta t, \Delta \kappa, \frac{\partial \bar{u}_i}{\partial x_j}, C_0, S_0$
\State Generate $n_p$ random unit vectors $\mathbf{e}$
\State Generate $n_k$ $\kappa$-values within $\Delta \kappa$
\State Compute initial energy spectrum: $ E^*=q^2A\kappa^{*4}\exp{\frac{-2\kappa^{*2}}{\kappa_p^2}}$
\State Generate initial spectra:
\State $\Phi_{ij}^*=\frac{E^*}{4\pi \kappa^{*2}}\left(\delta_{ij}-e^*_ie^*_j\right)$,  $\beta_i^*=0$, $\Gamma^*=\frac{E_d^*}{-2\pi \kappa^{*2}a_0^2\gamma}\left(1-e^{*2}_3\right)$
\For{$t=1:n_t$} 
\For{$k=1:n_k$}
\For{$n=1:n_p$}
\State Compute RHS for second moments at each particle for each $\kappa$ magnitude: 
\State $\frac{D}{Dt}(\Phi^*_{ij})=-\frac{\partial \bar{u}_i}{\partial x_k}\Phi^*_{kj}-\frac{\partial \bar{u}_j}{\partial x_k}\Phi^*_{ki}+\kappa e^*_ia^2\beta^*_j+\kappa e^*_ja^2\beta^*_i$
\State $\frac{D}{Dt}(\beta^*_i)=-\kappa e^*_k\Phi^*_{ki}-\frac{\partial \bar{u}_i}{\partial x_k}\beta^*_{k}+\kappa e^*_ia^2\Gamma^*$
\State $\frac{D}{Dt}(e^*_i)=-e^*_k\frac{\partial \bar{u}_k}{\partial x_i}+\frac{\partial \bar{u}_k}{\partial x_r}e^*_ke^*_re^*_i$
\State$\frac{D}{Dt}(\Gamma^*)=
-2\kappa e^*_i\beta^*_i$
\EndFor
\EndFor
\State update mean states:
\State $a=a_0\sqrt{\frac{1+C_0\Delta t\times t}{(1+C_0t)^{\gamma}}}$,
$\frac{\partial \bar{u}_1}{\partial x_1}=\frac{C_0}{1+C_0\Delta t\times t}$, $\frac{\partial \bar{u}_1}{\partial x_2}=\frac{S_0}{1+C_0\Delta t\times t}$
\State Compute next time step states with RK4
\EndFor
\State Integrate wavevector magnitude using trapezoidal or quadrature: $\Phi^{\nearrow \kappa*}_{ij}=\int \Phi^*_{ij}\kappa^2d\kappa$
\State Compute statistics with Monte Carlo Integration: $R_{ij}=4\pi\sum^{n_{p}}_{n=1}\Phi^{\nearrow \kappa*}_{ij}/n_{p}$
\end{algorithmic}
\end{algorithm}

\subsection{Axial compression}
\begin{figure}
\centering
\includegraphics[width=0.7\linewidth, trim = 0mm 0mm 0mm 0mm]{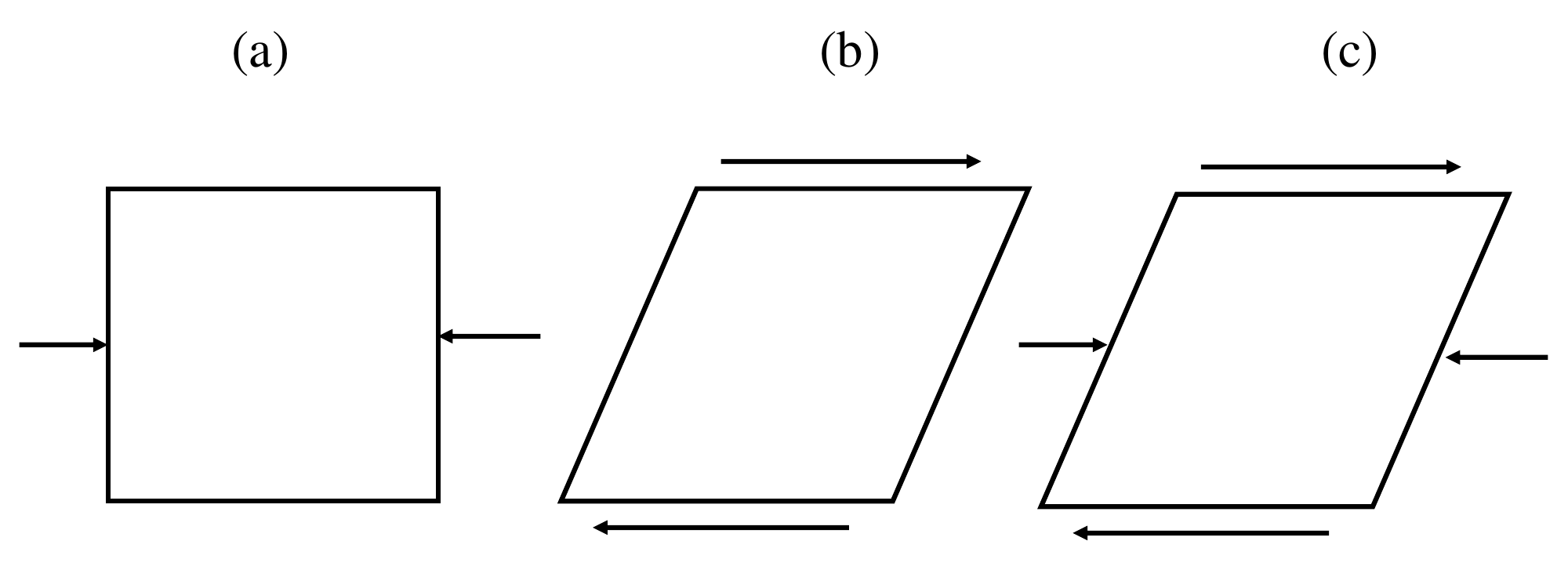}
\caption{Schematic of the mean velocity deformations for (a) axial compression ; (b) pure shear; (c) sheared compression.}
\label{fig:def}
\end{figure}

\begin{figure}
  \centering
  \includegraphics[width=0.55\linewidth, trim = 0mm 0mm 0mm 0mm]{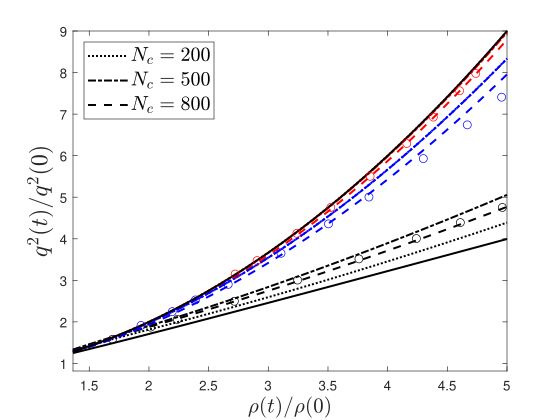}
  \caption{Histories turbulent kinetic energy amplification for three cases of varying distortion Mach number in axial compression: $M_{d0}=5$, black; $M_{d0}=29$, blue; $M_{d0}=87$, red; DNS of \cite{cambon1993rapid}, dots. Multiple predictions are shown for varying particle cluster counts varying from 200 to 800 clusters.}
\label{fig:axc_q}
\end{figure}

\begin{figure}
  \centering 
  \includegraphics[width=0.55\linewidth, trim = 0mm 0mm 0mm 0mm]{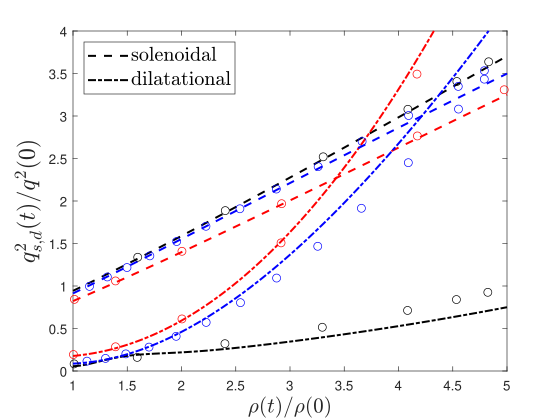}
  \caption{Evolution of solenoidal and dilatational components of turbulent kinetic energy amplification for three cases of varying distortion Mach number in in axial compression: $M_{d0}=5$, black; $M_{d0}=29$, blue; $M_{d0}=87$, red; DNS of \cite{cambon1993rapid}, dots.}
\label{fig:axc_qsd}
\end{figure}

The axial compression considered in this case is defined by
\begin{equation}
    G_{ij}=C\delta_{i1}\delta_{j1}, \; C(t)=\frac{C_0}{1+C_0t}=\frac{C_0\rho}{\rho_0},
\end{equation}
where the compression axis is in the $x_1$-direction (as shown in figure \ref{fig:def} case (a)), and $C$ represents the compression rate. Due to the change in volume/density (varying from $\rho/\rho_0=1$ to $\rho/\rho_0=5$), the speed of sound also changes. The relationship between the speed of sound and density ratio is derived from isentropic relations (\cite{mahesh1996interaction}, \cite{jacquin1993turbulence}), 
\begin{equation}
    a=a_0\sqrt{\frac{1+C_0 t}{(1+C_0t)^\gamma}}.
\end{equation}
As mentioned previously, the distortion Mach number $M_d$  is equal to the ratio between the initial acoustic timescale
and the mean distortion or compression timescale. So if $M_d<<1$, the acoustic timescale is much larger than the mean distortion timescale and a pure-acoustic/solenoidal regime is obtained \citep{durbin1992rapid}. In this solenoidal regime, the dilatational mode is balanced by the pressure gradient. The pressure is found with the pressure Poisson equation and consequentially describes the dilatational modes \citep{simone1997effect}. The energy is bounded in this regime and limits are well established (\citet{durbin1992rapid},\cite{simone1997effect}, \citet{jacquin1993turbulence}). On the other hand, if $M_d>>1$, the dilatational mode is no longer constrained by the pressure (hence the name pressure-released) and is fed directly by compression. For the $M_d>>1$ pressure-released limit, the turbulent kinetic energy amplification can be expressed as \citep{jacquin1993turbulence}
\begin{equation}
    \frac{q^2(t)}{q^2(0)}=\frac{2+(\rho(t)/\rho(0))^2}{3}.
\end{equation}
For the solenoidal limit $M_d<<1$, turbulent kinetic energy amplification is also found analytically using \citep{simone1997effect}
\begin{equation}
    \frac{q^2(t)}{q^2(0)}=\frac{1}{2}\left(1+(\rho(t)/\rho(0))^2\frac{\tan^{-1}\left(\sqrt{(\rho(t)/\rho(0))^2-1}\right)}{\sqrt{(\rho(t)/\rho(0))^2-1}}\right).
\end{equation}
Results for the axial compression cases are compared to DNS data of \citet{cambon1993rapid}. The initial conditions used for the three intermediate cases are listed in table \ref{tab:axc_initial_conditions}. Note that the distortion Mach number is defined with the negative compression rate for this problem, $M_d=-M_tCq^2/\epsilon$.
\begin{table}
  \begin{center}
\def~{\hphantom{0}}
  \begin{tabular}{lccccc}
      Case & $M_{t0}$  & $M_{d0}$   &   $-(Cq^2/\epsilon)_0$ & $Re_{t0}$ & $(q_d^2/q^2)_0$ \\[3pt]
      A & 0.025   & 5.0 & 194 & 358& 0.06\\
      B & 0.11   & 87 & 800 & 184&0.18\\
      C & 0.29  & 29 & 100 & 500&0.09\\
  \end{tabular}
  \caption{Initial conditions for axial compression cases.}
  \label{tab:axc_initial_conditions}
  \end{center}
\end{table}
  Unless otherwise stated, results presented use 800 particles and 60 discretizations of the wavevector magnitude for a range of $10^{-3}$ to $10^3$. The trapezoidal integration with respect to magnitude was used alongside Monte Carlo integration with respect to orientation. Figure \ref{fig:axc_q} shows the histories of the turbulent kinetic energy amplification for cases A-C with different cluster amounts, varying from 200 to 800. With increasing cluster count, predictions tend to improve and agree closer with the DNS data. The optimal cluster count varies based on the distortion Mach number. We recommend a starting point of 500 clusters for most cases. Figure \ref{fig:axc_q} also shows that increased compressibility ($M_d$) leads to an increase in the kinetic energy amplification. Both solenoidal and dilatational contributions to the kinetic energy are shown in figure \ref{fig:axc_qsd} and compared to DNS results. The solenoidal mode is nearly unaffected by the increased compression, which corroborates with the theory in \citet{cambon1993rapid}. A strong increase of the dilatational energy at the end of the compression is also captured by CPRM.

\begin{figure}
  \centering \includegraphics[width=0.55\linewidth, trim = 0mm 0mm 0mm 0mm]{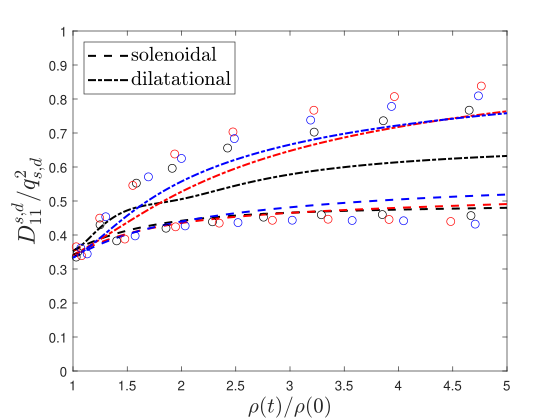}
  \caption{Evolution of solenoidal and dilatational components of the $D_{11}/q^2$ structure tensor component for three cases of varying distortion Mach number in axial compression: $M_{d0}=5$, black; $M_{d0}=29$, blue; $M_{d0}=87$, red; DNS of \cite{cambon1993rapid}, dots.}
\label{fig:axc_D}
\end{figure}

The dimensionality structure tensor $D_{11}$ component is also computed and compared to DNS data. Figure \ref{fig:axc_D} shows the histories of normalized solenoidal and dilatational components of $D_{11}$ and matches the trends of the solenoidal/dilatational components of the DNS data. Note that for axial compression, $D^d_{11}=R_{11}^d$ \citep{cambon1993rapid}. The dimensionality provides information on the angular distribution of the spectral energy. In the pressure-released case, the alignment of the wavevector with the compression direction increases $D_{11}/q^2$ towards one, while the angular distribution of the spectral energy remains unchanged. Meanwhile, in the solenoidal limit, the angular distribution of spectral energy in the plane normal to the compression direction opposes the wavevector motion so that a slower positive increase of $D_{11}/q^2$ is displayed. The $M_{d0}=5$ case has poor $D_{11}^d/q^2_d$ prediction, showing the CPRM does not capture this spectral distribution opposing the wavevector motion. Reasons for this disagreement include wavevector discretization errors and the resampling technique inaccuracy for the wavevector magnitude, which remains constant due to the infinite integration. On the contrast, when evolving the wavevector magnitude, the dimensionality results are over-predicted, but the case near the solenoidal limit was much closer.

While the CPRM is expected to compare well with RDT, the results show some discrepancies between the high-fidelity data. One reason for this is that the comparing data in this section is from the DNS of \citet{cambon1993rapid}. We do not expect RDT to match exactly with DNS, and thus small disagreement is reasonable. \citet{cambon1993rapid} also stated that the solenoidal ratio of $D_{11}/q^2$ predicted by the DNS was found to be slightly lower than the RDT analytical prediction. This finding is also supported by the results for the solenoidal CPRM $D_{11}/q^2$ predictions. Finally, CPRM requires different information for initialization than what is given for DNS. The pressure-velocity spectrum $\beta_i$ was initialized to zero for this case, while the pressure spectrum $\Gamma$ was initialized with the dilatational energy. These initializations are approximate as the energy spectrum for the pressure-velocity and pressure spectra are not known. Therefore, since the evolution of the turbulence statistics depend on initial conditions, the inexact initial conditions contribute to the error seen in the results.

It is worth noting the differences and analogies of the current CPRM formulation to the traditional RDT used by \cite{jacquin1993turbulence}. CPRM uses the same set of governing equations and assumptions as \cite{jacquin1993turbulence}, but ends up with a different and reduced set of evolution equations. This is because CPRM only used deforming coordinates/Lagrangian perspective as the coordinate transformation whereas the RDT of \cite{jacquin1993turbulence} used deforming coordinates and a Craya-Herring transformation. CPRM also solves for the full-field variables and does not require the Helmholtz decomposition as compared to the RDT which was developed using this decomposition. The normalized pressure variable $\hat{\wp}$ removes the explicit appearance of i in the evolution equation and so CPRM can be solved without any need to handle imaginary components, whereas the simplified RDT equations presented by \cite{jacquin1993turbulence} did have an explicit imaginary number appear in the pressure terms for the general  (equations 30 and 32) but not in the reduced equations for axial-compression case. The last difference is the method of numerical integration. Both CPRM and RDT of \cite{jacquin1993turbulence} require numerical integration, but CPRM is solved using stochastic methods while the numerical integration method is not specified by \cite{jacquin1993turbulence}, but likely used some form of fast-Fourier transforms.

\subsubsection{Axial Compression: A surrogate for Shock-turbulence interaction}
\begin{figure}
  \centering
  \includegraphics[width=0.55\linewidth, trim = 0mm 0mm 0mm 0mm]{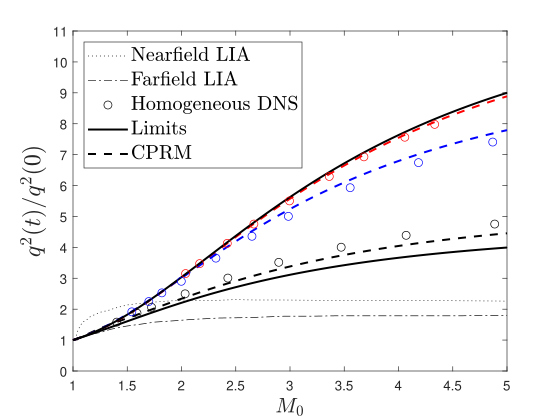}
  \caption{Evolution of turbulent kinetic energy amplification for three intermediate case in axial compression compared to LIA and limiting RDT: $M_{d0}=5$, black; $M_{d0}=29$, blue; $M_{d0}=87$, red.}
\label{fig:axc_shock}
\end{figure}

Axial compression can be used as a simplified model for shock-turbulence interaction. \citet{jacquin1993turbulence} used this model problem to compare solenoidal and pressure released limits against the linear interaction approximation (LIA) theory of \citet{ribner1954shock}. Homogeneous axially-compressed turbulence is only a model of shock-turbulence interaction because true shock-turbulence interaction does not meet the quasi-homogeneous requirement that $\ell<<\delta$ where $\delta$ is the mean scale (the thickness of the shock). For further evaluation of CPRM, we follow the canonical problem of \citet{jacquin1993turbulence}, where an initial flow at Mach number $M_0$ impinges a shock of strength $\Delta \bar{u}/\bar{u}_0=(\rho-\rho_0)/\rho$, such that the density ratio is given by
\begin{equation}
    \frac{\rho}{\rho_0}=\frac{(\gamma +1)M_0^2}{2+(\gamma-1)M_0^2}.
\end{equation}
Figure \ref{fig:axc_shock} shows the turbulent kinetic energy amplification for varying $M_0$. The disagreement with LIA (computed by \citet{jacquin1993turbulence}) is significant, and follows the same results obtained by \citet{jacquin1993turbulence}. CPRM performance in figure \ref{fig:axc_shock} is identical to figure \ref{fig:axc_q} because the problem dynamics are the same, just with increased compression rate $C$ (where increasing $C$ approaches the pressure-released solution). CPRM does match farfield LIA at very short times/low $M_0$, but does not match the nearfield LIA predictions.

\subsection{Pure shear}
The deformation for pure shear is shown in case (b) of figure \ref{fig:def}. Pure shear is defined as 
\begin{equation}
    G_{ij}=S\delta_{i1}\delta_{j2}, \; S(t)=S_0,
\end{equation}
where $S_0$ is the initial shear rate, which remains constant. No volumetric compression is experienced ($\rho/\rho_0=1$) and thus the speed of sound and mean pressure/density are constant in time $a=a_0$, $\bar{p}(t)=\bar{p}_0$, $\bar{\rho}(t)=\bar{\rho}$.
For anisotropies $b_{ij}=R_{ij}/q^2$, the pressure-released limit is found by neglecting the pressure terms in the evolution equation for $\Phi^{*}_{ij}$. The evolution of the turbulent kinetic energy amplification in this limit is, however, found analytically when the initial turbulence conditions are isotropic \citep{simone1997effect},
\begin{equation} \label{eq:ps_q}
    \frac{q^2(t)}{q^2(0)}=1+\frac{(St)^2}{3}.
\end{equation}
Simple analytical solutions only exist for the pressure-released limit. Solenoidal limits are evaluated by decreasing the distortion Mach number $M_d<<1$ and using the solenoidal form of the evolution equations for $\Phi^s_{ij}$. Cluster count varied from 500 near the solenoidal limit to 700 near the pressure-released limit. A 20-point trapezoidal integration was used for the wavevector magnitude. Due to instabilities at high $St$ values, the evolution of the wavevector magnitude was taken into account in the evolution equations for the spectra (no resampling used). A very fine time step was required at high $St$ values, reaching as low as $\Delta t/t_f=0.001$.

\begin{figure}
\centering
\includegraphics[width=0.55\linewidth, trim = 0mm 0mm 0mm 0mm]{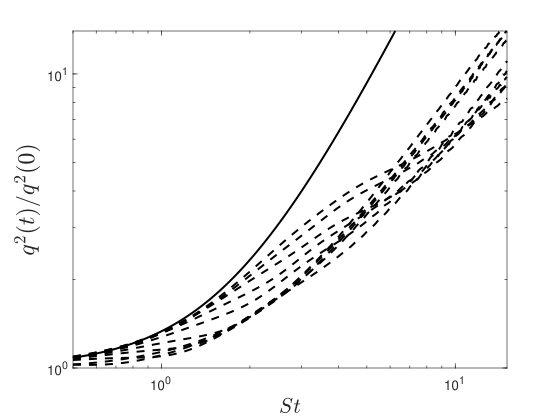}
\caption{Turbulent kinetic energy amplification evolution in pure shear for varying distortion Mach numbers prescribed in table \ref{tab:ps_initial_conditions}. Solid line given by equation \ref{eq:ps_q}. Results comparable with \citet{bertsch2012rapid} and \citet{lavin2012flow}.}
\label{fig:ps_q}
\end{figure}

Evolution of the turbulent kinetic energy amplification is presented in figure \ref{fig:ps_q}. As expected, the turbulent kinetic energy amplification is bounded by the pressure-released limit and increases with initial $M_d$. Amplification increase is much greater than that of the DNS data from \citet{simone1997effect}. The early-time monotonic increase is predicted by CPRM but the later-time linear behavior does not match the DNS. This disagreement is likely due to the re-meshing technique used in the DNS that resulted in loss of turbulent kinetic energy and thus lower growth rates. However, CPRM does predict the three-stage evolution of the TKE amplification as described by \cite{bertsch2012rapid}, \cite{lavin2012flow}, and \cite{kumar2014stabilizing}. The intermediate flattening of the turbulent kinetic energy evolution at $St \approx 4-6$ seen in figure \ref{fig:ps_q}. In the first stage, irrespective of the initial modal Mach number $M_m=S/a|\boldsymbol{\kappa}|$, rapid growth of the turbulent kinetic energy is experienced and matches the rate of the pressure-released solution. This stage is defined by $St/\sqrt{M_m(0)}<1$, where the fluctuating pressure change is too slow to influence the velocity field, which grows unbounded. Transition from the first to the second stage begins when the pressure starts to influence the velocity field. Through this influence, the kinetic energy evolution begins to depart from the pressure-released solution. We see a sudden reduction/stabilization in the kinetic energy amplification growth in figure \ref{fig:ps_q} around $St\approx 1.5$. The flattening of kinetic energy is bounded by $1<M_m(t)<\sqrt{M_m(0)}$. The third stage of kinetic energy evolution begins where the kinetic energy grows rapidly again. This occurs when the acoustic time scale $a_0t|\boldsymbol{\kappa_0}|$ becomes smaller than the mean shear time scale $St$. This occurs at approximately $a_0t|\boldsymbol{\kappa_0}| \approx 3$. At this stage, the dilatational component of the flow is small and thus the turbulence is predominately governed by the solenoidal dynamics.


\begin{figure}
\centering
\hspace*{0.4cm} \begin{subfigure}{0.7\textwidth}
\centering
\includegraphics[width=\linewidth, trim = 0mm 0mm 0mm 0mm]{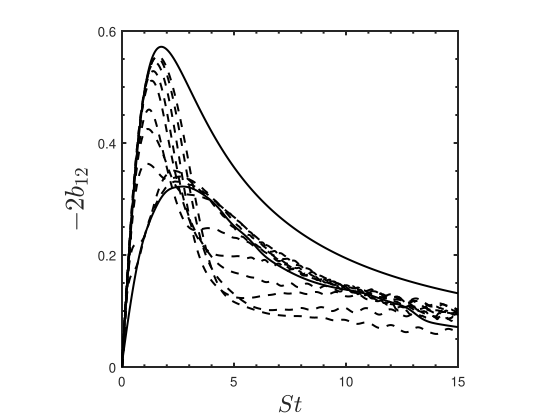}
\caption{CPRM shear anisotropy histories }
\end{subfigure} \\
\begin{subfigure}{0.55\textwidth}
\centering
\includegraphics[width=\linewidth, trim = 0mm 0mm 0mm 0mm]{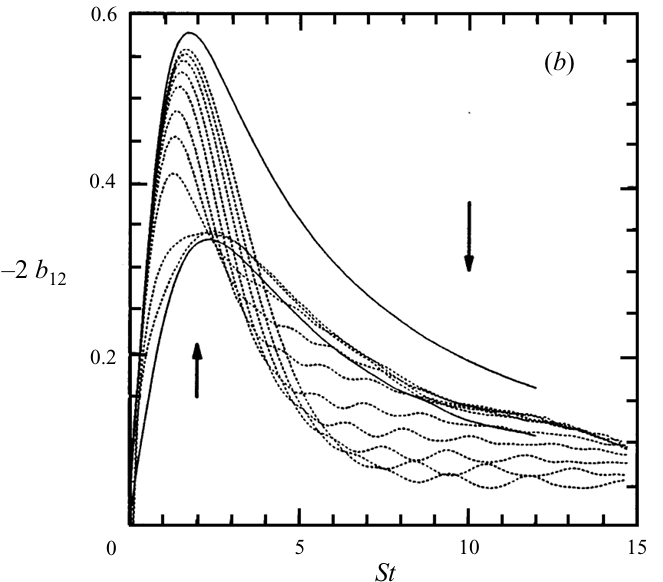}
\caption{RDT shear anisotropy histories of \citet{simone1997effect}}
\end{subfigure}
\caption{Comparison of shear anisotropy evolution in pure shear between CPRM (a) and RDT (b) for varying distortion Mach numbers prescribed in table \ref{tab:ps_initial_conditions}. Top solid line is the pressure-released limit while lower solid line is the solenoidal limit.}
\label{fig:ps_b12}
\end{figure}

Another important feature of compressible pure shear is the ``crossover point" displayed in the evolution of the shear anisotropy component $b_{12}$. A crossover point at $St\approx4$, where trends with $M_d$ reverse, is predicted by CPRM in figure \ref{fig:ps_b12}. Before the crossover, $b_{12}$ increases with $M_d$ but decreases soon afterwards. The rate of decrease depends on the distortion Mach number $M_d$. The crossover occurs due to coupling between the solenoidal and dilatational modes \cite{simone1997effect}.

\begin{figure}
\centering
\includegraphics[width=0.55\linewidth, trim = 0mm 0mm 0mm 0mm]{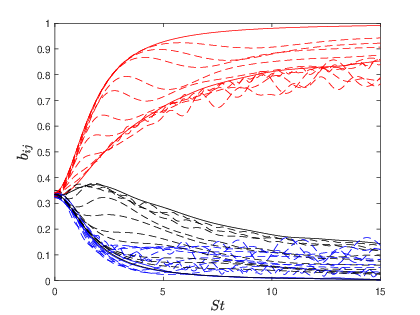}
\caption{Evolution of normal anisotropy components in pure shear for varying distortion Mach numbers prescribed in table \ref{tab:ps_initial_conditions}: pressure-released/solenoidal limits, top/bottom solid lines; $b_{11}$, red; $b_{22}$, blue; $b_{33}$, black.}
\label{fig:ps_bii}
\end{figure}

Predictions in figure \ref{fig:ps_b12} do not match as well with the RDT data of \citet{simone1997effect} at large $St$. It is noted in \citet{simone1997effect} that the ``waviness" at large times in the RDT histories is the result of the wavenumber discretization and resolution at large wavevectors. Since the integration with respect to wavevector magnitude is done numerically, this also impacts the large time solution of CPRM and is likely the cause for the mismatch between CPRM and RDT of \citet{simone1997effect}. The histories of the normal $b_{ij}$ components are shown in figure \ref{fig:ps_bii}. RDT/DNS data for these quantities are not available, so we simply compare to the pressure-released and solenoidal limits. For most of the range, the predictions stay bounded. Recall that analytical solutions for pressure-released and solenoidal anisotropy limits do not exists, so they are approximated by using large/small $M_d$ values with pressure-terms ignored or using the solenoidal model. This approximation is a reason why it does not always bound the lower/higher $M_d$ cases. The oscillation due to wavenumber discretization is also pronounced and causes the long-time solution to go past the limits. 

\begin{figure}
\centering
\includegraphics[width=0.55\linewidth, trim = 0mm 0mm 0mm 0mm]{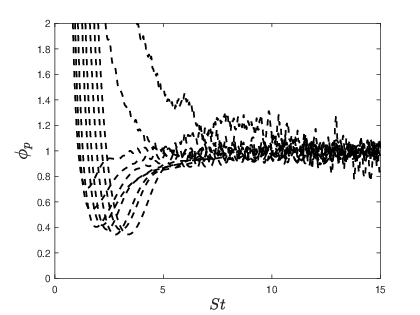}
\caption{Evolution of the equipartition variable $\phi_p$ for varying higher distortion Mach numbers prescribed in table \ref{tab:ps_initial_conditions}. Relaxation to equipartition is obtained when TKE curve reaches second stage}
\label{fig:equipartition}
\end{figure}

The equipartition between the pressure fluctuations and dilatational component of the turbulent kinetic energy is also recorded at large mean shear time scales in figure \ref{fig:equipartition}. The equipartition variable is defined as
\begin{equation}
    \phi_p=\frac{\overline{u'_2u'_2}\bar{p}\gamma\bar{\rho}}{\overline{p'p'}}=\iiint^{\infty}_{-\infty}\frac{\Phi_{22}}{a^2\Gamma}d^3\boldsymbol{\kappa}.
\end{equation}
\cite{bertsch2012rapid} used this definition of equipartition between the pressure and dilatational velocity components because the flow-normal kinetic energy component is nearly equal to the dilatational kinetic energy of the flow field at late times due to the alignment of the wavenumber vector with the flow-normal direction in homogeneous shear. The evolution of the equipartition as predicted by CPRM shows a similar sharp decay and oscillatory behavior near $\phi_p=1$ as the RDT analysis done by \cite{bertsch2012rapid}. Since the pressure-dilatation is a driving mechanism for distribution of energy between the mechanical and internal modes, it is expected that the equipartition follows a similar trend to the three-stages of the pressure correlation/turbulent kinetic energy evolution as described previously. Indeed, it is seen that the start of the plateau of turbulent kinetic energy coincides with the equipartition of energy between dilatation and pressure modes ($St\approx 6$). The evolution of the pressure statistics $\overline{p'p'}/\bar{p}^2\gamma^2$ are also verified by comparing to the RDT data by \cite{lavin2012flow}.

\begin{figure}
\centering
\includegraphics[width=0.55\linewidth, trim = 0mm 0mm 0mm 0mm]{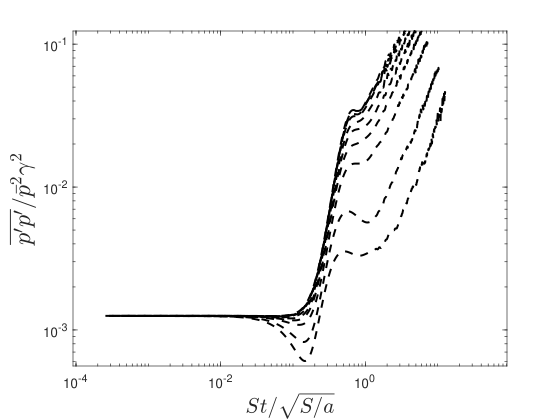}
\caption{Evolution of the normalized pressure statistics $\overline{p'p'}/\bar{p}^2\gamma^2$ for cases A-J in table \ref{tab:ps_initial_conditions}. Data matches the trends seen in \cite{lavin2012flow}.}
\label{fig:ps_pressure}
\end{figure}

\begin{figure}
\centering\includegraphics[width=0.55\linewidth, trim = 0mm 0mm 0mm 0mm]{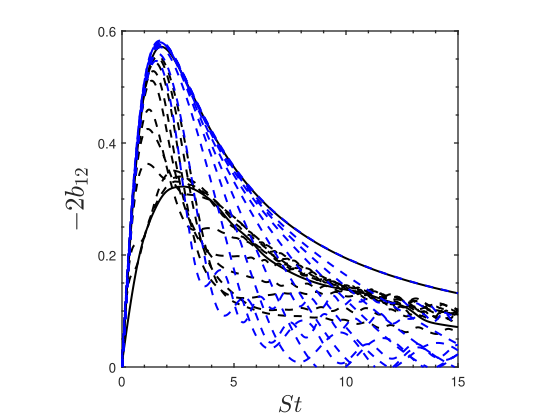}
\caption{Evolution of shear anisotropy components in pure shear for varying distortion Mach numbers with the new CPRM (black) compared to the model by \citet{yu2007extension} (blue).}
\label{fig:ps_compare}
\end{figure}

We compare our model results with that of \citet{yu2007extension}. For this comparison, the histories of $b_{12}$ are compared for the same range of $M_d$ given in table \ref{tab:ps_initial_conditions}. From figure \ref{fig:ps_compare}, it is clear that the model by \citet{yu2007extension} does predict the correct cross-over point, but the small details like the decay afterwards or the magnitude of the cross-over point peaks do not match the RDT data as well as CPRM does. This is likely due to the fact that \citet{yu2007extension} did not account for the appearance of the wavevector magnitude in the evolution equations which requires one to evolve the velocity spectrum tensor rather than the directional velocity spectrum tensor. The appearance of the $\kappa^3$ factor in the pressure terms after conversion to spherical coordinates renders it impossible to analytically factor out the directional velocity spectrum tensor. The new model also addresses a discrepancy in the past compressible RDT model where the definitions for second moments in Fourier space considered the entire complex expression,  which does not align with the symmetric velocity spectrum tensor definition by \citet{durbin2011statistical} used to compute the Reynolds stress. This resulted in complex evolution equations where real and imaginary components needed to be treated. 

While CPRM does not match the RDT data exactly, its predictions are much improved compared to past models. Again, due to the formulation of CPRM and the requirement of the $\beta_i$ and $\Gamma$ spectra, the initialization is not exactly the same as the RDT of \citet{simone1997effect} since we do not know the pressure-velocity and pressure energy spectra. This is one reason why the predictions do not match the RDT results exactly, along with possible discretization errors. Note that CPRM and the RDT of \cite{simone1997effect} are still very similar methods, sharing the same differences and analogies as the comparisons made between CPRM and the RDT of \cite{jacquin1993turbulence} discussed in the previous section.
\begin{table}
  \begin{center}
\def~{\hphantom{0}}
  \begin{tabular}{lcccc}
      Case & $M_{t0}$  & $M_{d0}$   &   $(Sq^2/\epsilon)_0$ & $Re_{t0}$ \\[3pt]
      A & 0.25   & 2.7 & 10.7 & 296\\
      B & 0.25   & 4.0 & 16.0 & 296\\
      C & 0.25  & 8.3 & 33.1 & 296\\
      D & 0.25   & 12.0 & 48.0 & 296\\
      E & 0.25 & 16.5 & 66.2 & 296\\
      F & 0.25 & 24.0 & 96.1 & 296\\
      G & 0.25 & 32.0 & 128.1 & 296\\
      H & 0.25 & 42.7 & 170.8 & 296\\
      I & 0.25 & 53.4 & 213.5 & 296\\
      J & 0.25 & 66.7 & 266.9 & 296\\
  \end{tabular}
  \caption{homogeneous isotropic turbulence initial conditions for pure-shear cases.}
  \label{tab:ps_initial_conditions}
  \end{center}
\end{table}

\begin{figure}
\centering
\begin{subfigure}{0.45\textwidth}
\centering
\includegraphics[width=\linewidth, trim = 0mm 0mm 0mm 0mm]{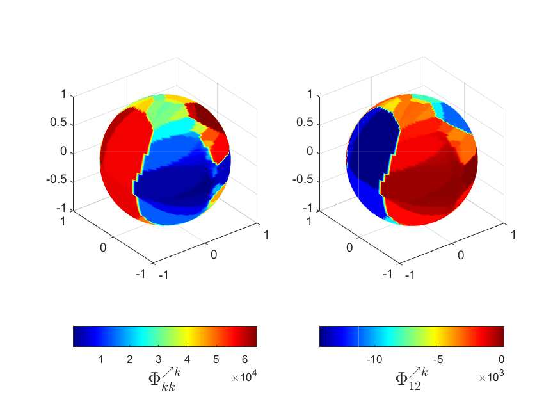}
\caption{Distrbutions for case A}
\end{subfigure} 
\begin{subfigure}{0.45\textwidth}
\centering
\includegraphics[width=\linewidth, trim = 0mm 0mm 0mm 0mm]{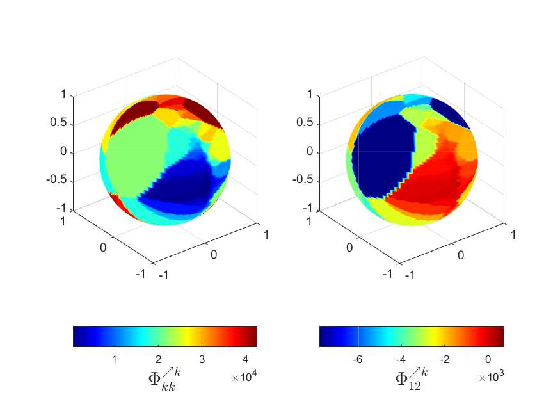}
\caption{Distrbutions for case D}
\end{subfigure} 
\begin{subfigure}{0.45\textwidth}
\centering
\includegraphics[width=\linewidth, trim = 0mm 0mm 0mm 0mm]{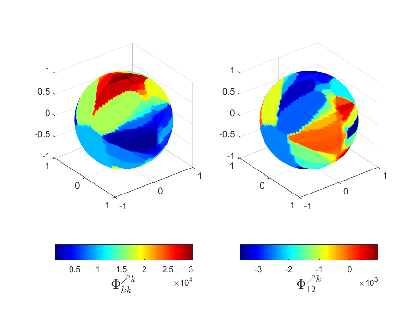}
\caption{Distrbutions for case G}
\end{subfigure} 
\begin{subfigure}{0.45\textwidth}
\centering
\includegraphics[width=\linewidth, trim = 0mm 0mm 0mm 0mm]{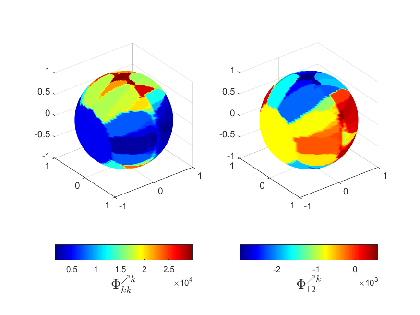}
\caption{Distrbutions for case J}
\end{subfigure} 
\caption{Visualization of the directional velocity spectrum trace and shear component in pure shear for varying distortion Mach numbers prescribed in case A,D,G, and J. Results are shown at time $St=10$}
\label{fig:ps_spheres}
\end{figure}

The distribution of the directional velocity spectrum over the unit-$\boldsymbol{\kappa}$ sphere is visualized in figure \ref{fig:ps_spheres}. These distributions evolve from a homogeneous isotropic (uniform) distribution over the unit sphere. The magnitudes of $\Phi^{\nearrow \kappa*}_{kk}$ and $\Phi^{\nearrow \kappa*}_{12}$ tend to decrease with increasing $M_d$. This matches the behavior seen in figure \ref{fig:ps_b12}. Another interesting feature is that the distribution pattern of $\Phi^{\nearrow \kappa*}_{kk}$ and $\Phi^{\nearrow \kappa*}_{12}$ are identical for the same case and only vary in magnitude range.

\subsection{Sheared compression}
\begin{figure}
  \centering
  \includegraphics[width=0.55\linewidth, trim = 0mm 0mm 0mm 0mm]{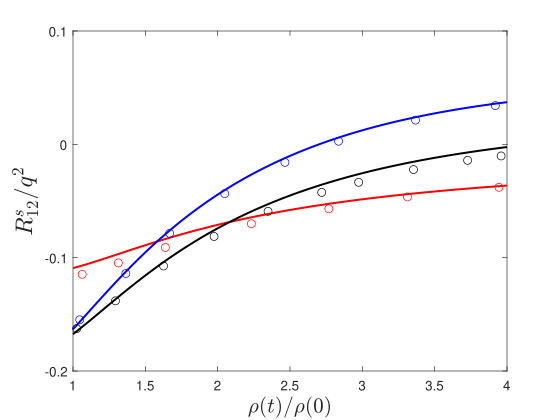}
  \caption{Evolution of solenoidal component of $b_{12}$ for varying anisotropic initial condition in sheared compression; blue: $\mathcal{S}_0=1$, black: $\mathcal{S}_0=2$, red: $\mathcal{S}_0=3$, dots: RDT of \citet{mahesh1996interaction}.}
\label{fig:sc_b12}
\end{figure}

Sheared compression is a combination of the last two cases. It is defined by
\begin{equation}
    G_{ij}=C\delta_{i1}\delta_{j1}+S\delta_{i1}\delta_{j2}, \; C(t)=\frac{C_0}{1+C_0t},\; S(t)=\frac{S_0}{1+C_0t}.
\end{equation}
The shear and compression rates change with time to satisfy the compressible Euler equations. To match the results of \cite{mahesh1996interaction}, anisotropic initial spectra for the three second-moments are required. This is done by applying pure shear for $t_{ps}$ time, defined by the non-dimensional time $\mathcal{S}_0=S_0t_{ps}$. Note that for this problem, only solenoidal turbulence is considered. Thus the solenoidal model is used to evolve the directional velocity spectrum tensor $\Phi^{\nearrow \kappa*}_{ij}$. The initial shear is intended to generate the anisotropic field but is negligible compared to the applied compression i.e. $S_0/C_0<<1$. Thus a low shear-compression ratio of $S_0/C_0=0.1$ is used \citep{mahesh1996interaction}. Only $\mathcal{S}_0$ varies across the different cases, so all other initial variables remain the same. For this case, $M_{t0}=0.025$, $Re_{t0}=358$, and $M_{d0}=0.5$. Again, 800 clusters are used, and wavevector magnitude integration is not required since the directional spectrum is evolved. The histories of $b_{12}$ are plotted in figure \ref{fig:sc_b12}.
Each curve has different initial total shear, which increases from red to blue. Agreement with the RDT of \citet{mahesh1996interaction} shows that the CPRM solenoidal method  works well for anisotropic initial conditions. Slight discrepancies are likely due to imperfect initialization, noting that the worst results also has the worst initialization (start off with the most error). The initial anisotropic velocity spectrum was not given by \citet{mahesh1996interaction} and thus CPRM was used to obtain the initial spectrum by running the pure shear case for $\mathcal{S}_0$ total shear. Therefore, we are unable to match the initial spectrum exactly to that of \citet{mahesh1996interaction}. Figure \ref{fig:sc_b12} displays important trends of decreasing shear anisotropy and change of sign for the case with sufficiently large total volumetric strain (which correlates with larger initial total shear). 

\begin{figure}
\centering
\includegraphics[width=0.55\linewidth, trim = 0mm 0mm 0mm 0mm]{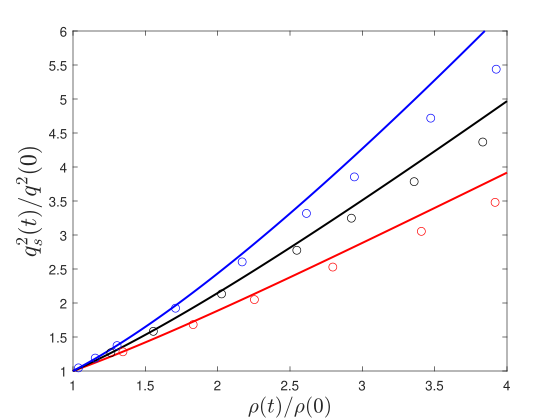}
  \caption{Evolution of solenoidal component of turbulent kinetic energy for varying anisotropic initial condition in sheared compression; blue: $\mathcal{S}_0=1$, black: $\mathcal{S}_0=2$, red: $\mathcal{S}_0=3$, dots: RDT of \citet{mahesh1996interaction}.}
\label{fig:sc_q}
\end{figure}

\begin{figure}
  \centering
  \includegraphics[width=0.55\linewidth, trim = 0mm 0mm 0mm 0mm]{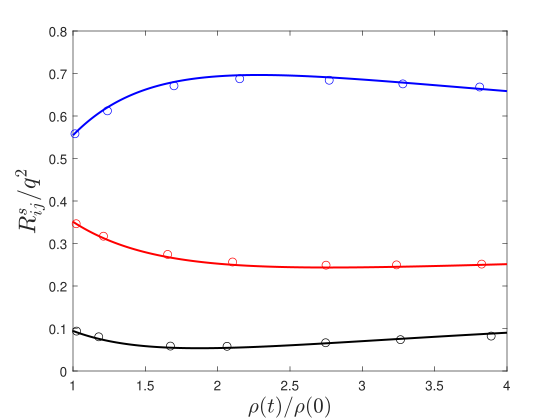}
  \caption{Evolution of normal components of $b_{ij}$ in sheared compression for anisotropic initial condition $\mathcal{S}_0=3$: $b_{11}$, blue; $b_{22}$, black; $b_{33}$, red; RDT of \citet{mahesh1996interaction}, dots.}
\label{fig:sc_bii}
\end{figure}

Figure \ref{fig:sc_q} shows the turbulent kinetic energy amplification for the same cases. Again, due to inexact initialization, the turbulent kinetic energy amplification is also not exactly predicted but still has the same increasing trends for the three different total shear values. Figure \ref{fig:sc_bii} shows the normal components of $b_{ij}$ compared against the RDT of \citet{mahesh1996interaction} and shows nearly exact agreement, where the compression increases the contribution of 
$u_1'$ to the turbulent kinetic energy and decreases that of $u'_2$ and $u_{3}'$. Only results for the maximum initial total shear $\mathcal{S}_0=3$ are given in figure \ref{fig:sc_bii}.

\subsection{Plane strain and axisymmetric contraction}
CPRM is applicable to any generic mean deformations that are rapid enough such that RDT still applies. To demonstrate this, CPRM is applied to plane strain and axisymmetric deformations, defined by
\begin{equation}
    \setlength{\arraycolsep}{0pt}
\renewcommand{\arraystretch}{1.3}
\frac{\partial \bar{u}_i}{\partial x_j}_{ps} = \left[
\begin{array}{ccc}
  S  &  \ \ 0  \ \  &  0   \\
  0  & \ \  -S \ \  &   0  \\
  0  & \ \  0 \ \  &   0  \\
\end{array}  \right],\;
\frac{\partial \bar{u}_i}{\partial x_j}_{axc} = \left[
\begin{array}{ccc}
  S & \ \  0 \ \ &  0   \\
  0  & \ \ -S/2 \ \  &   0  \\
  0  & \ \ 0  &  \ \ -S/2  \\
\end{array}  \right].
\end{equation}
For this problem, DNS/RDT data sets are not available and thus we cannot determine the model's accuracy. Instead, we simply verify that the model is performing adequately by comparing predictions to the approximate solenoidal and pressure-released limits. For all cases, 800 clusters were used along with a 20-weight Gauss-Laguerre quadrature.
\begin{figure}
  \centering
  \includegraphics[width=0.55\linewidth, trim = 0mm 0mm 0mm 0mm]{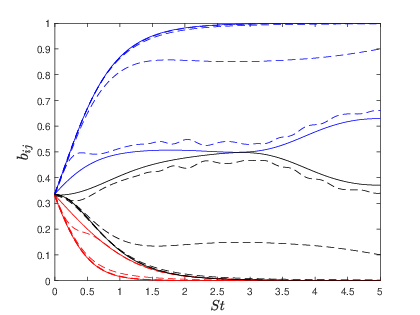}
  \caption{Evolution of normal components of $b_{ij}$ subject to plane strain for initial conditions A,D,G, and J listed in table \ref{tab:ps_initial_conditions}. Same legend as in figure \ref{fig:ps_bii}.}
\label{fig:plst_bii}
\end{figure}
\begin{figure}
  \centering
  \includegraphics[width=0.55\linewidth, trim = 0mm 0mm 0mm 0mm]{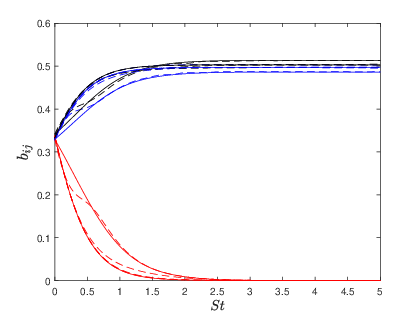}
  \caption{Evolution of normal components of $b_{ij}$ subject to axisymmetric contraction for initial conditions A,D,G, and J listed in table \ref{tab:ps_initial_conditions}. Same legend as in figure \ref{fig:ps_bii}.}
\label{fig:ac_bii}
\end{figure}
Figure \ref{fig:plst_bii} shows the normal $b_{ij}$ components for plane strain, while figure \ref{fig:ac_bii} shows $b_{ij}$ for axisymmetric contraction. The results fall within the limits, showing that the the qualitative behavior of CPRM is correct. It is important to note that while cases A and J are near the limits, they are not so small/large that we expect it to follow the limiting-lines exactly. The results show that the model can be applied to generic rapid deformations, although more high-fidelity data is needed to evaluate the accuracy of CPRM.

\section{Concluding remarks}
\label{sec:Concluding remarks}

We developed a new stochastic wave vector model for rapidly-distorted compressible turbulence subject to the restrictions of homogeneity and small density fluctuations. By representing the evolution equations in spectral space and evolving the pressure-velocity and pressure spectra $\beta_i$, $\Gamma$, the velocity spectrum tensor evolution equation was closed without any modeling assumptions. This preserves the exact nature of RDT and only requires approximation of the spectral-to-physical transformation integrals. A new variable called the normalized pressure fluctuation $\wp$ enabled the formulation of $\beta_i$ and $\Gamma$, and simplified the second-moment evolution equations by removing explicit imaginary terms. 

Motivation for a stochastic representation was given by considering the use of  Monte Carlo integration  to transform to physical space. The stochastic model uses stochastic samples or virtual particles to represent realizations of the random variables $u^*_i$, $\kappa^*_i$, and $\wp^*$, which were then used to compute the second-moment statistics $\Phi_{ij}$, $\beta_i$, and $\Gamma$.  Consistency between the Fourier and stochastic representation was shown by comparing the Eulerian velocity spectrum evolution equations derived from the linearized stochastic PDF equations and traditional RDT. The idea of sample/particle clustering with respect to wavevector orientation was also motivated and described, where it became advantageous when a non-stochastic method of integration for the infinite integral of wavevector magnitude was used. By working in spectral space, the computation of higher-order tensors was also possible due to the availability of the wavevector information. These higher-order tensors provide more information on the turbulent field and can be used to develop improved models for non-linear or inhomogeneous problems. 

Next, the new model was applied to three homogeneous compressible problems: axially-compressed stationary and shear turbulence, and pure shear deformations. Results were generally in agreement with the available DNS/RDT data. The axial compression case was discussed as an idealized shock-turbulence interaction model, where dilatational and solenoidal turbulent kinetic energy amplification histories were shown to be in agreement with DNS data. Evolution of the  turbulent kinetic energy amplification was also evaluated for different cluster amounts and showed a converging trend towards  true DNS solutions. The pure shear case demonstrated the applicability and performance of CPRM across a range of distortion Mach numbers $M_d$ with improvement in the prediction of the shear anisotropy evolution compared to the compressible RDT model of \citet{yu2007extension}. Equipartition of pressure fluctuations and dilatational velocity fluctuations was also observed for large times, along with the three phases of turbulent kinetic energy evolution. The sheared compression case showed that the model is able to predict stresses/amplification accurately when an anisotropic initial shear is axially compressed. All three cases, however, were found to be sensitive to the initial conditions, which could not be matched exactly. 

Finally, since the formulation of the model made no assumption on the form of the deformation tensor, it can be applied to any generic rapid mean deformation. Plane strain and axisymmetric contraction deformations were applied to verify that the resulting Reynolds stress anisotropy histories for four different distortion Mach numbers were bounded within the solenoidal (small $M_{d0}$) and pressure-released (large $M_{d0}$) limits. Initialization methods were discussed and developed for the velocity spectrum tensor. 

We emphasize that this work is merely a starting point for development of practical turbulence models that maintain  consistency in  the rapid-distortion regime. The exact limiting behavior overcomes the limitations of eddy-viscosity and other standard assumptions, and enables the computation of higher-order tensors to bring in more information on the physics of turbulence. Extension to general deformation and inhomogeneities are the next steps towards development of CPRM for use in more complex problems. There are several  ways of accounting for nonlinear effects in the current formulation. A limiting model can be created similar to the nonlinear incompressible PRM of \cite{kassinos1996particle}, or more sophisticated approaches can be used to directly model the multi-point triadic interactions such as Eddy-damped Quasi-normal Markovian models \citep{orszag1970analytical}. Application of the stochastic wavevector formulation to inhomogeneous compressible turbulence can be accomplished in two ways and will be explored in future work. Although the exact equations presented in equations \ref{eq:fm_vel}-\ref{eq:presSpecEvo} cannot be used directly in inhomogeneous cases, due to the requirement of periodicity for Fourier bases, a similar set of equations can be derived by using a non-periodic basis such as Chebyshev polynomials \citep{canutospectral}. This will require a new derivation of the basic formulation, but the use of a spectral representation should still allow exact closure of the linearized equations since the differentiation properties of the Chebyshev polynomials are similar to the Fourier basis (spatial derivatives become point-wise wavevector-basis coefficient products). Alterately, CPRM can be extended to inhomogeneous flows using a quasi-homogeneous approach similar to the elliptic-relaxation technique developed by \cite{durbin1991near}. This approach is more direct in that it represents inhomogeneities without requiring the original tensor evolution equations to be modified. This approach was used by \cite{kassinos2000structure} to extend the incompressible PRM to wall-bounded flows.


\backsection[Funding]{Noah Zambrano was supported by the National Science Foundation Graduate Research Fellowship Program. Karthik Duraisamy was supported by the OUSD(RE) Grant \# N00014-21-1-295.}

\backsection[Declaration of interests]{The authors report no conflict of interest.}




\appendix
\section{Solenoidal and Dilatational Decomposition}
 The solenoidal-dilatational velocity decomposition $\hat{u}_i=\hat{u}_i^s+\hat{u}_i^d$ is used to define the solenoidal, dilatational, and solenoidal-dilatational cross-correlations,
\begin{equation}\label{eq:vel-spec-sol}
\Phi^s_{ij} =\frac{1}{2}
\left(\overline{ \hat{u}^s_i\hat{u}^{s+}_j}+\overline{ \hat{u}^{s+}_i\hat{u}^s_j}\right),\;\Phi^{d}_{ij} =\frac{1}{2}
\left(\overline{ \hat{u}^d_i\hat{u}^{d+}_j}+\overline{ \hat{u}^{d+}_i\hat{u}^d_j}\right),\;\Phi^{sd}_{ij} =\frac{1}{2}
\left(\overline{ \hat{u}^s_i\hat{u}^{d+}_j}+\overline{ \hat{u}^{s+}_i\hat{u}^d_j}\right),\end{equation}
\begin{equation} \beta^s_{i}=\frac{1}{2}\left(\overline{\hat{\wp}^+\hat{u}^s_i} + \overline{\hat{\wp}\hat{u}_i^{s+}}\right),\; \beta^d_{i}=\frac{1}{2}\left(\overline{\hat{\wp}^+\hat{u}^d_i} + \overline{\hat{\wp}\hat{u}_i^{d+}}\right).\end{equation}
Note that $\Phi^{sd}_{ij}\neq\Phi^{ds}_{ij}$ and, in general, $\Phi^{sd}_{ij}$ is not symmetric (unlike $\Phi^s_{ij}$ and $\Phi^d_{ij}$). The velocity spectrum solenoidal/dilatational components are related by 
\begin{equation}
\Phi_{ij}=\Phi^s_{ij}+\Phi^d_{ij}+\Phi^{sd}_{ij}+\Phi^{ds}_{ij}.
\end{equation} 
When turbulence is solenoidal, there is no need to introduce $\hat{\wp}$ because the imaginary pressure term from equation \ref{eq:fluc-vel} is removed by substituting in the pressure-Poisson equation \ref{eq:pressure_poisson}. The dilatational velocity evolution is determined once $D\hat{u}_i/Dt$ and $D\hat{u}_i^s/Dt$ are known.
\begin{equation}\label{eq:fluc-vel_sol}\frac{D\hat{u}^s_i}{Dt}=-\frac{\partial \bar{u}_i}{\partial x_k}\hat{u}^s_k+2\hat{u}^s_j\frac{\kappa_i\kappa_k}{\kappa^2}\frac{\partial \bar{u}_k}{\partial x_j},\end{equation}
\begin{equation}\label{eq:fluc-vel_dil}\frac{D\hat{u}^d_i}{Dt}=-\frac{\partial \bar{u}_i}{\partial x_k}\hat{u}^d_k+\kappa_ia^2\hat{\wp}-2\hat{u}^s_j\frac{\kappa_i\kappa_k}{\kappa^2}\frac{\partial \bar{u}_k}{\partial x_j},\end{equation}
The second-moment evolution equations for the solenoidal, dilatational, and cross solenoidal-dilatational velocity spectra are now derived as
\begin{equation} \label{eq:velSpecTenEvo_sol}
\frac{D}{Dt}(\Phi_{ij}^s)=-\frac{\partial \bar{u}_i}{\partial x_k}\Phi_{kj}^s-\frac{\partial \bar{u}_j}{\partial x_k}\Phi_{ki}^s+2\frac{\partial \bar{u}_k}{\partial x_m}\left(\Phi_{im}^s\frac{\kappa_k\kappa_j}{\kappa^2}+\Phi_{jm}^s\frac{\kappa_k\kappa_i}{\kappa^2}\right), \end{equation}
\begin{equation} \label{eq:velSpecTenEvo_dil}
\frac{D}{Dt}(\Phi_{ij}^d)=-\frac{\partial \bar{u}_i}{\partial x_k}\Phi_{kj}^d-\frac{\partial \bar{u}_j}{\partial x_k}\Phi_{ki}^d+a^2\left(\kappa_i\beta^d_j+\kappa_j\beta^d_i\right)-2\frac{\partial \bar{u}_k}{\partial x_m}\frac{\kappa_k}{\kappa^2}\left(\Phi_{mi}^{sd}\kappa_j+\Phi_{mj}^{sd}\kappa_i\right), \end{equation}
\begin{equation} \label{eq:velSpecTenEvo_cross}
\frac{D}{Dt}(\Phi_{ij}^{sd})=-\frac{\partial \bar{u}_i}{\partial x_k}\Phi_{kj}^{sd}-\frac{\partial \bar{u}_j}{\partial x_k}\Phi_{ik}^{sd}+\kappa_ja^2\beta^s_i+2\frac{\partial \bar{u}_k}{\partial x_m}\frac{\kappa_k}{\kappa^2}\left(\Phi_{mj}^{sd}\kappa_i-\Phi_{mi}^{s}\kappa_j\right), \end{equation}
\begin{equation}\frac{D}{Dt}(\beta^s_i)=-\kappa_k\Phi^s_{ki}-\kappa_k\Phi^{sd}_{ik}-\frac{\partial \bar{u}_i}{\partial x_k}\beta^s_{k}+2\beta^s_j\frac{\partial \bar{u}_k}{\partial x_j}\frac{\kappa_i\kappa_k}{\kappa^2},\end{equation}
\begin{equation}\frac{D}{Dt}(\beta^d_i)=-\kappa_k\Phi^{sd}_{ki}-\kappa_k\Phi^{d}_{ki}-\frac{\partial \bar{u}_i}{\partial x_k}\beta^d_{k}-2\beta^s_j\frac{\partial \bar{u}_k}{\partial x_j}\frac{\kappa_i\kappa_k}{\kappa^2}+\kappa_ia^2\Gamma.\end{equation}
Note that in the solenoidal velocity spectrum tensor, the dependence on wavevector magnitude is removed. The solenoidal velocity spectrum tensor evolution equation is identical to the evolution equation in the incompressible formulation of \citet{kassinos1995structure}. This is because the divergence-free property of the solenoidal modes allows the use of the pressure-Poisson equation, the mean compression effects on the fluctuating density are neglected, and the mean density does not vary in space (to be homogeneous). This reinforces the observation in other studies \citep{cambon1993rapid} where the mean compressibility has little to no effect on the solenoidal velocity components. 

Now, the initialization of a turbulence field that contains both solenoidal and dilatational components is described. Since equation \ref{eq:init-sol} initializes the solenoidal field, the dilatational field must be initialized separately. If the solenoidal-dilatational cross correlation $\Phi^{sd}_{ij}$ is assumed to be zero, the relation $\Phi_{ij}\approx\Phi^s_{ij}+\Phi^d_{ij}, q^2=q^2_s+q^2_d$ is satisfied. Since $q_s^2/q_d^2$ is specified at the start, then the dilatational field is found using the same method as described by \citet{simone1997effect},
\begin{equation}\Phi^d_{ij}(\boldsymbol{\kappa}^n)=\frac{E_d(\kappa^n)}{2\pi (\kappa^n)^2}.
\end{equation}
\bibliographystyle{jfm}
\bibliography{jfm}

\end{document}